\documentclass[aps,preprintnumbers,showpacs,twocolumn,amsmath,amssymb]{revtex4}
\usepackage{graphicx}
\usepackage{feynmp}
\usepackage{multirow}
\usepackage{hyperref}
\usepackage{epstopdf}


\begin{document}
\title{A $T$-Matrix Calculation for in-Medium Heavy-Quark Gluon Scattering}

\author{K.~Huggins and R.~Rapp}
\affiliation{$^{1}$Cyclotron Institute and Department of Physics \& Astronomy, 
Texas A\&M University, College Station TX 77843-3366}
\date{\today}

\begin{abstract}
The interactions of charm and bottom quarks in a Quark-Gluon 
Plasma (QGP) are evaluated using a thermodynamic 2-body $T$-matrix. 
We specifically focus on heavy-quark (HQ) interactions with thermal gluons 
with an input potential motivated by lattice-QCD computations of the 
HQ free energy. The latter is implemented into a field-theoretic ansatz
for color-Coulomb and (remnants of) confining interactions. This, 
in particular, enables to discuss corrections to the potential approach,
specifically hard-thermal-loop corrections to the vertices, relativistic 
corrections deduced from pertinent Feynman diagrams, and a suitable 
projection on transverse thermal gluons. The resulting potentials 
are applied to compute scattering amplitudes in different color channels 
and utilized for a calculation
of the corresponding HQ drag coefficient in the QGP. A factor of $\sim$2-3 
enhancement over perturbative results is obtained, mainly driven by the 
resummation in the attractive color-channels.  
\end{abstract}


\maketitle
\section{Introduction}
The description of the Quark-Gluon Plasma (QGP) requires an understanding 
of how its constituents propagate in the heat bath. The information on
pertinent transport properties can ultimately be used to understand
how the dense matter produced in heavy-ion colliders such as RHIC and LHC 
evolves. A starting point of such investigations is the basic two-body
interaction between constituents of the QGP. The identification of an 
in-medium force is a formidable task, especially in a nonperturbative 
regime as is likely present in the QGP close to the critical temperature, 
$T_c$, and up to 2-3 times its value. For example, lattice-QCD (lQCD) 
calculations of the heavy-quark (HQ) free energy have found that 
contributions from the confining force, i.e., linear in the separation, 
$r$, between heavy quark ($Q$) and anti-quark ($\bar Q$), persist up to 
well above $T_c$~\cite{Pisarski:2002ji,Kaczmarek:2005ui,Petreczky:2004}. 
The HQ sector 
is particularly suitable to study the basic QCD two-body force since a 
large quark mass renders elastic interactions with spacelike exchange 
kinematics dominant.  This leads to a potential interaction which 
facilitates the reduction of a pertinent scattering equation from the 
4-dimensional Bethe-Salpeter one to a 3-dimensional 
Lippmann-Schwinger~\cite{PhysRevC.38.51}
form. As emphasized in Ref.~\cite{Rapp:2009my}, the spacelike exchange 
kinematics prevalent in HQ bound states also applies to the scattering 
of a single heavy quark, thus opening the possibility for a comprehensive 
treatment of quarkonia and HQ transport in the QGP~\cite{Riek:2010fk}. 
This has recently been carried out in a thermodynamic $T$-matrix approach 
where the input potential is resummed in ladder approximation and medium 
effects are incorporated in both the interaction kernel and the 
intermediate particle propagation. Thus far, the interactions of heavy 
quarks in the QGP have mostly been considered with light 
quarks~\cite{vanHees:2007me,Riek:2010fk}, which could be done in a 
fairly straightforward extension of 
quarkonia~\cite{Cabrera:2006wh,PhysRevC.72.064905}.

In the present work we extend the $T$-matrix approach to HQ-gluon 
scattering. Compared to the heavy-light quark case, this requires to 
revisit the relativistic corrections to the potential vertices and 
associated subtleties due to the different dimension of the scattering
amplitude, as well as the appropriate transverse polarization of the in- 
and outgoing thermal gluons. Once suitable amendments are deduced and
implemented we compute the pertinent $T$-matrices in all available color 
projections and obtain thermal relaxation rates as defined by an underlying 
Fokker-Planck equation. In particular, we will compare our results to 
perturbative evaluations in order to quantify the relevance 
of the confining and higher-order terms in the $T$-matrix.

Our article is organized as follows. In Sec.~\ref{sec_pot} we briefly
recall the field-theoretical model used to implement the lQCD-motivated 
static potentials. In Sec.~\ref{sec_rel} we discuss the relativistic 
corrections to the color-Coulomb and confining terms of the potential, 
with special care of the different dimension of the quark-gluon relative 
to the quark-quark $T$-matrix and a restriction to transverse degrees of 
the physical gluons. In Sec.~\ref{sec_tmat} we calculate the HQ-gluon 
$T$-matrices in different color channels and elucidate the role of the 
confining and the higher-order rescattering terms. In 
Sec.~\ref{sec_trans} we utilize the HQ-gluon $T$-matrices to evaluate 
the HQ friction coefficient at different temperatures as a function of 
3-momentum, decomposed into its color and angular-momentum contributions. 
We summarize and conclude in Sec.~\ref{sec_concl}.

\section{Construction of the in-Medium Potential}
\label{sec_pot}
\begin{figure*}[!t]
\includegraphics[scale=.5]{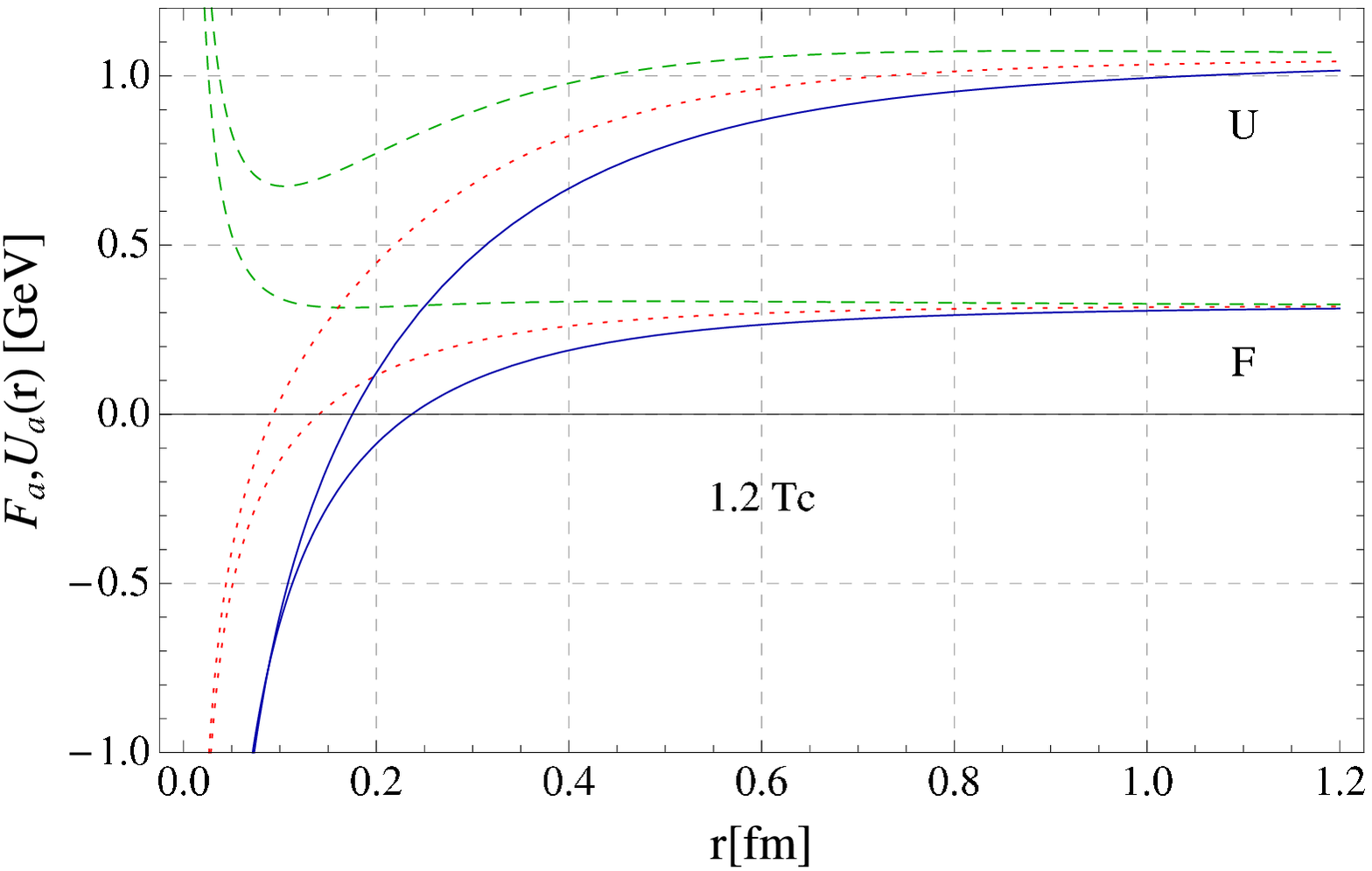}
\includegraphics[scale=.5]{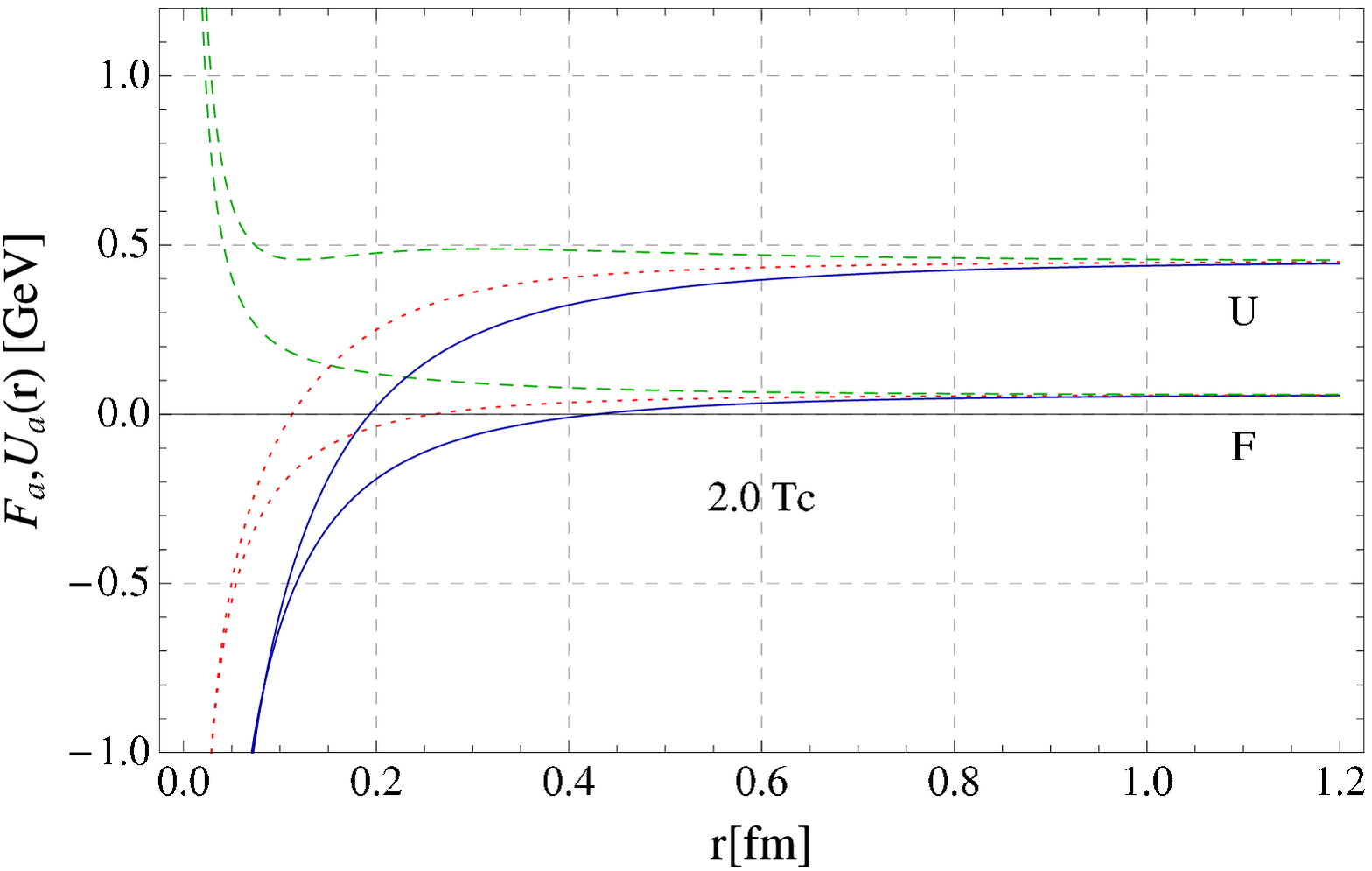}
\includegraphics[scale=.5]{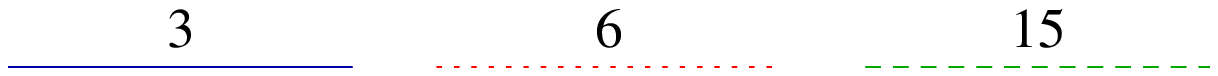}
\caption{In-medium HQ free and internal energies in the triplet (solid 
lines), sextet (dotted lines), and 15-plet (dashed lines) channels at
temperatures $T=1.2\,T_c$ (left) and 2\,$T_c$ (right). 
(Color online.)}
\label{figpot}
\end{figure*}
Recent computations of the static HQ free energy in thermal 
lQCD~\cite{Bazavov:2009us} have revived potential-based approaches to 
quarkonia in the QGP. Since a functional fit to static coordinate-space 
quantities is somewhat limited in flexibility for microscopic 
applications, we follow Ref.~\cite{Riek:2010fk} where a microscopic 
model~\cite{Megias:2007pq,Megias:2005ve} for color-Coulomb and confining 
terms has been adopted. A fit has been performed to the color-average free 
energies at different temperatures (from 1.2\,$T_c$ to $\sim$2\,$T_c$) 
using 4 model parameters characterizing the overall strength ($\alpha_s$, 
$m_G$) and screening properties ($m_D$, $\tilde{m}_D$) of each of the 
two contributions (here, we focus on the lQCD data of 
Ref.~\cite{Kaczmarek:2007pb}, denoted as ``potential-1" in 
Ref.~\cite{Riek:2010fk}; another set of lQCD data studied therein
leads to slightly stronger effects).  
The different color projections are then obtained as
\begin{eqnarray}
F_a(T,r)&=&
-\frac{4}{3}\alpha_s(\frac{C_a}{r}e^{-m_Dr}+\frac{m_G^2}{2\tilde{m}_D}e^{-\tilde{m}_Dr}
\nonumber\\
&&-\frac{m_G^2}{2\tilde{m}_D}+m_D)\ ,
\label{F}
\end{eqnarray}
where it is assumed that the string interaction (second term), as well as 
the long-distance limit represented by the last two terms, are color-blind.
For the color-Coulomb interaction (first term), one assumes the standard 
(perturbative) 
Casimir scaling resulting in the following coefficients for quark-antiquark, 
quark-quark and quark-gluon channels, respectively~\cite{PhysRevD.70.054507},
\begin{eqnarray}
\nonumber 
C_1&=&1   \qquad , \  C_8=-1/8  \ ,  
\\
\nonumber C_{6}&=&-1/4  \ , \ C_{\bar{3}}=1/2   \ \ \ ,
\\
C_3&=&9/8 \quad \, , \  C_6=3/8  \ \ \ , \  C_{15}=-3/8 \ .
\end{eqnarray}
The internal energy follows as
\begin{equation}
U(T,r)=F(T,r)-T\frac{d}{dT}F(T,r)  \ .
\end{equation}
The resulting coordinate-space potentials in the HQ-gluon color projections
are summarized in Fig.~\ref{figpot}.

There is currently no consensus as to whether the free or internal energy (or
combinations thereof) should be used as a static potential in a Schr\"odinger 
(or Lippmann-Schwinger) equation. We will perform our calculations for both $U$ 
and $F$. As usual, we will subtract off the infinite-distance limit to define
the genuine two-body interaction contribution in each case, 
\begin{equation}
V_{a}(T,r)=X_a(T,r)-X(T,\infty)\hspace{.06in},
\hspace{.06in}X=F \text{ or } U \ ,  
\label{V_a}
\end{equation}
and reinsert $X(T,\infty)$ into the calculation by interpreting it as a
temperature-dependent mass term of the heavy anti-/quarks.
For a more reliable application at higher energies, it remains to elaborate 
the effects of relativistic corrections for the case at hand, i.e., HQ-gluon 
scattering.  

\section{Relativistic Corrections to Potential Terms}
\label{sec_rel}
\begin{figure}[!b]
\begin{center}
\unitlength = 1mm
\begin{fmffile}{Diagrams/tchannelgluon}
\begin{fmfgraph*}(45,30)
\fmfleft{i1,i2}
\fmfright{o1,o2}

\fmflabel{$p_1$}{i1}
\fmflabel{$p_2$}{i2}
\fmflabel{$p_3$}{o1}
\fmflabel{$p_4$}{o2}

\fmf{fermion}{i1,v1,o1}
\fmf{curly}{i2,v2,o2}
\fmf{curly}{v1,v2}
\end{fmfgraph*}
\end{fmffile}
\end{center}
\caption{Feynman diagram for $Q$-$g$ scattering via $t$-channel gluon 
exchange. In the center-of-mass system, $q=(p_2-p_1)/2$ and $q'=(p_4-p_3)/2$. }
\label{fig:t-channel}
\end{figure}
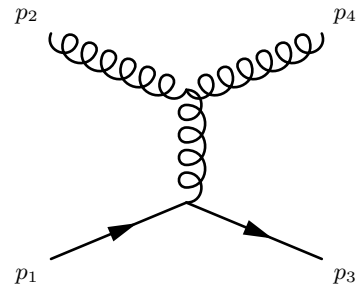

To identify relativistic corrections to the nonrelativistic potential we
will be analyzing Feynman diagrams associated 
with the Coulomb and string terms, $V_c$ and $V_s$, respectively. 
For the Coulombic part, we refer to the $t$-channel exchange diagram for 
quark-gluon scattering displayed in Fig.~\ref{fig:t-channel} (Compton
scattering diagrams with an intermediate HQ propagator are parametrically
suppressed by a power of the HQ mass). 
To establish a relationship between the fully relativistic scattering 
amplitude, $\mathcal{M}$, and the nonrelativistic $T$-matrix, $T$ 
(following from the Fourier transform of the static coordinate-space 
potential), we write down pertinent expressions of the cross
section. For fermion-boson scattering with Bjorken-Drell conventions
(where the Dirac spinors are normalized as $\bar u u =1$),
the former reads \cite{bjorken1964relativistic}
\begin{eqnarray}
\nonumber d\sigma&=&\frac{(2\pi)^4m_1m_3}{\omega_1\omega_2|\bold{v}_1-\bold{v}_2|}
\int \frac{d^3p_3}{2\omega_3(2\pi)^3} \frac{d^3p_4}{2\omega_4(2\pi)^3}
\\
&&\times|\mathcal{M}|^2\delta^{(4)}(p_3+p_4-p_1-p_2) \ ;
\label{sig_M}
\end{eqnarray}
note that in this convention $\mathcal{M}$ carries the dimension of 
inverse energy, dim$(\mathcal{M})=E^{-1}$ (for fermion-fermion scattering, 
dim$(\mathcal{M})=E^{-2}$, see also Ref.~\cite{Riek:2010fk}; this is
convenient for taking the nonrelativistic limit to recover
the nonrelativistic $T$-matrix).  
In terms of the nonrelativistic $T$-matrix, the cross section can
be expressed as~\cite{goldberger2004collision} 
\begin{eqnarray}
 d\sigma&=&\frac{(2\pi)^4}{\omega_1\omega_2|\bold{v}_1-\bold{v}_2|}
\int\frac{d^3p_3}{2\omega_3 (2\pi)^3}\frac{d^3p_4}{2\omega_4(2\pi)^3}
\\
\nonumber&&\times |\sqrt{\omega_1\omega_2}T\sqrt{2\omega_32\omega_4}|^2
\delta^{(4)}(p_3+p_4-p_1-p_2)\ .
\label{sig_T}
\end{eqnarray}
Comparison with Eq.~(\ref{sig_M}) thus leads to 
\begin{equation}
\sqrt{\omega_1\omega_2}T\sqrt{2\omega_32\omega_4}=\sqrt{m_1m_3}\mathcal{M}\ .
\label{T-M}
\end{equation}
In Ref.~\cite{Herrmann:1992} the relation between the nonrelativistic
$T$-matrix and a properly generalized relativistic form, $\tilde{T}$,
has been worked out for the fermion-fermion case as
\begin{align}
T=\sqrt{\frac{m_1m_2m_3m_4}{\omega_1\omega_2\omega_3\omega_4}}\tilde{T} \ ,
\label{Ttilde}
\end{align}
which is the same result obtained in Ref.~\cite{Riek:2010fk}, i.e.,
$\tilde{T}$ is the relevant quantity for our purposes here. However, since
we are interested in fermion-boson scattering, an additional dimensionful
normalization factor appears. This can be obtained by inserting the
expression for $\tilde{T}$ from Eq.~(\ref{Ttilde}) into Eq.~(\ref{T-M}). 
One obtains
\begin{equation}
\tilde{T}=\frac{1}{\sqrt{2m_22m_4}}\mathcal{M} \ ,
\end{equation}
where we recall that $\mathcal{M}$ is the fully relativistic scattering
amplitude in Bjorken-Drell normalization. 

To proceed, we consider the Born approximation, where 
$\tilde{T}=\tilde{V}$, and evaluate $\mathcal{M}$ using Feynman 
diagrams from an underlying Lagrangian. 
As before, we split $\mathcal{M}$ into a Coulombic and string part, 
\begin{equation}
\mathcal{M}=\mathcal{M}_c+\mathcal{M}_s \ ,
\end{equation}
where the Coulombic part is given by the gluon-exchange diagram in 
Fig.~\ref{fig:t-channel} and the string contribution by 
Fig.~\ref{fig:scalar}. Let us first consider the Coulomb part which 
we factorize into a Yukawa-like propagator including coupling and 
color factors and a spinor-dependent quantity,
\begin{align}
\mathcal{M}_c=-i&\underbrace{\frac{C_A g^2}{t-m_D^2}}
\underbrace{\bar{u}(p_3)(-i\gamma^\mu) 
u(p_1)\epsilon(p_4)_{\rho}^*\Gamma^{\rho\lambda}_\mu\epsilon(p_2)_\lambda} \ ,
\nonumber \\
&\hspace{.04in}\text{Yukawa}\hspace{.84in}\text{Spinor}
\label{eqn:V}
\end{align}
where $u$ denote the HQ spinors, $\epsilon^\mu$ the gluon polarization 
vectors, $\Gamma^{\rho\lambda}_\mu$ the 3-gluon vertex, and 
$1/({t-m_D^2})$ the gluon-exchange propagator. The 3-gluon vertex is 
given by 
\begin{align}
\nonumber\Gamma^{\rho\lambda}_\mu=
&-(-g^{\rho\lambda}(p_4+p_2)_\mu+g^{\lambda}_\mu(p_2-q)^\rho
\\
&+g^{\rho}_\mu(q+p_4)^\lambda) \ .
\end{align}
The Yukawa part is merely the Fourier transform of the Coulombic 
static potential ansatz given in Eq.~(\ref{V_a}), while the second
factor encodes the relativistic corrections. That is,
\begin{equation}
\tilde{V}_c=\frac{V_c}{\sqrt{2m_22m_4}}|-i\bar{u}(p_3)(-i\gamma^\mu) 
u(p_1)\epsilon(p_4)_{\rho}^*\Gamma^{\rho\lambda}_\mu\epsilon(p_2)_\lambda|\ .
\label{eqn:Vrel}
\end{equation}
The spinor part is calculated by taking its square and then contracting 
across the vertices. This allows to read off the leading relativistic 
corrections to our static potential ansatz (a similar procedure will be
carried out for the string term, cf.~Eq.~(\ref{Lstring}) and below). 

\begin{figure}
\unitlength = 1mm
\begin{fmffile}{Diagrams/gluonloop}
\fmfframe(0,0)(0,7){
	\begin{fmfshrink}{.4}
	\begin{fmfgraph*}(45,30)
	\fmfleft{i1}
	\fmfright{o1,o2}
	\fmflabel{$p_1$}{i1}
	\fmflabel{$p_3$}{o1}
	\fmflabel{$p_2$}{o2}
	\fmf{curly,tension=3,width=.01}{i1,v1}
	\fmf{curly,tension=1}{v1,v3}
	\fmf{curly,tension=1}{v1,v2}
	\fmf{curly,tension=.1}{v3,v2}
	\fmf{curly,tension=3}{v2,o2}
	\fmf{curly,tension=3}{v3,o1}
	\end{fmfgraph*}
	\end{fmfshrink}
}
\end{fmffile}
\unitlength = 1mm
\begin{fmffile}{Diagrams/ghostloop}
\fmfframe(0,0)(0,7){
\begin{fmfshrink}{.4}
\begin{fmfgraph*}(45,30)
\fmfleft{i1}
\fmfright{o1,o2}
\fmflabel{$p_1$}{i1}
\fmflabel{$p_3$}{o1}
\fmflabel{$p_2$}{o2}
\fmf{curly,tension=3,width=.01}{i1,v1}
\fmf{dashes,tension=1}{v1,v3}
\fmf{dashes,tension=1}{v1,v2}
\fmf{curly,tension=.1}{v3,v2}
\fmf{curly,tension=3}{v2,o2}
\fmf{curly,tension=3}{v3,o1}
\end{fmfgraph*}
\end{fmfshrink}
}
\end{fmffile}


\unitlength = 1mm
\begin{fmffile}{Diagrams/quarkloop}
\fmfframe(0,0)(0,7){
\begin{fmfshrink}{.4}
\begin{fmfgraph*}(45,30)
\fmfleft{i1}
\fmfright{o1,o2}
\fmflabel{$p_1$}{i1}
\fmflabel{$p_3$}{o1}
\fmflabel{$p_2$}{o2}
\fmf{curly,tension=3,width=.01}{i1,v1}
\fmf{plain,tension=1}{v1,v3}
\fmf{plain,tension=1}{v1,v2}
\fmf{plain,tension=.1}{v3,v2}
\fmf{curly,tension=3}{v2,o2}
\fmf{curly,tension=3}{v3,o1}
\end{fmfgraph*}
\end{fmfshrink}
}
\end{fmffile}


\unitlength = 1mm
\begin{fmffile}{Diagrams/gluoncircleloop}

\begin{fmfshrink}{.4}
\begin{fmfgraph*}(45,30)
\fmfleft{i1}
\fmfright{o1}
\begin{fmfstraight}
\fmfbottom{b1,b2,b3}
\end{fmfstraight}
\fmftop{t1}
\fmflabel{$p_1$}{b1}
\fmflabel{$p_2$}{t1}
\fmflabel{$p_3$}{b3}
\fmf{curly}{b1,b2,b3}
\fmf{curly,left=1,tension=.3}{b2,v2,b2}
\fmf{curly}{v2,t1}
\end{fmfgraph*}
\end{fmfshrink}

\end{fmffile}
\caption{Contributions to the 3-gluon vertex corrections. 
Curly, solid and dashed lines represent gluons, quarks and ghosts,
respectively.}
\label{corrections}
\end{figure}
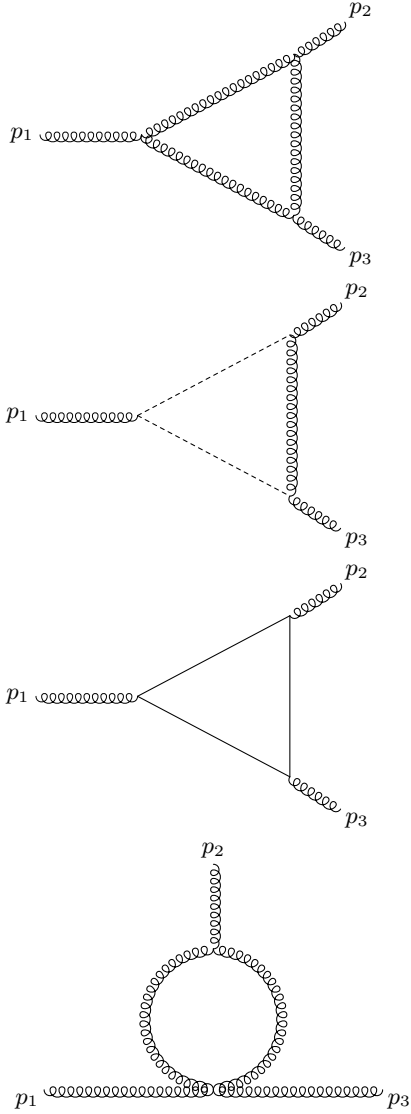

\begin{figure}[t]
\begin{center}
\unitlength = 1mm
\begin{fmffile}{Diagrams/scalar}
\begin{fmfgraph*}(45,30)
\fmfleft{i1,i2}
\fmfright{o1,o2}
\fmflabel{$p_1$}{i1}
\fmflabel{$p_2$}{i2}
\fmflabel{$p_3$}{o1}
\fmflabel{$p_4$}{o2}
\fmf{fermion}{i1,v1,o1}
\fmf{curly}{i2,v1,o2}
\end{fmfgraph*}
\end{fmffile}
\end{center}
\caption{Schematic approximation of the nonperturbative string term by a scalar 4-point interaction.}
\label{fig:scalar}
\end{figure}
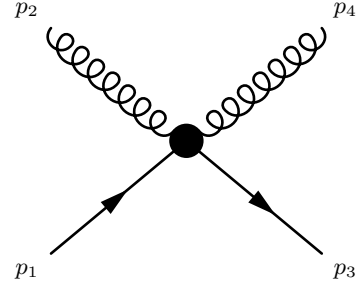

A consistent treatment at finite temperature requires the 
consideration of thermal vertex corrections which we analyze within the 
Hard Thermal Loop (HTL) framework. 
For the 3-gluon vertex, we have contributions from a gluon loop, 
ghost loop and quark loop as illustrated in Fig.~\ref{corrections}.
The gluon contribution reads
\begin{align}
\nonumber\Gamma^{(gl)}(p_1,p_2,p_3)=&ig^3C_Af_{abc}\int\frac{d^4k}{4\pi}9{k}_{\mu}{k}_{\nu}{k}_{\rho}\\
&\times\Delta(k)\Delta(p_2-k)\Delta(p_3+k) \  ,
\end{align}
where the $\Delta$'s indicate intermediate gluon propagators. Utilizing 
standard power counting techniques~\cite{Braaten:1990,Braaten1990310},
the propagator contributions will schematically integrate to
\begin{equation}
\int\frac{d^4k}{(2\pi)^4}\Delta(k)\Delta(p_2-k)\Delta(p_3+k)
\propto\int\frac{dk}{kp}X \ ,
\end{equation}
where $X$ depends on the quantum statistics of the interacting particles.  
There are three possible cases: (i) $X$ involves a sum of statistical 
factors of the same kind (both fermion or both boson), (ii) $X$ is a 
difference of statistical factors of the same kind, and 
(iii) $X$ is a difference of statistical factors of different kinds.  
Case (i) and (ii) occur for the 3-gluon vertex, while (iii) occurs 
for the quark-gluon vertex. However, case (i) will drop out because it 
is associated with a $T=0$ divergence, and the 3-gluon vertex is 
only linearly divergent. Examining the 3-gluon vertex (ii), we have, 
schematically, $X\propto\frac{dn}{dk}\rightarrow 1/T$.  Combining the 
above equation with the kinematic contributions of the vertex, the gluon 
HTL correction will behave as
\begin{equation}
g^3\int\frac{dk}{kp}k^3X\propto g^3T^2p^{-1} \ .
\end{equation}
Thus, for $p\sim gT$ the HTL correction is of order $g^2T$, which is 
the same as the zeroth order 3-gluon vertex. However, in the regime of
interest here, the gluons in the heat bath are hard and thus $p\approx T$.  
This renders HTLs subleading and it is sufficient to use the bare vertex 
at finite temperature. A similar argument is carried out for the 
quark-gluon vertex and the other loop corrections displayed in 
Fig.~\ref{corrections}. 

Let us now return to the task of analyzing the diagrammatic representation
of our potential. Toward this end we calculate the potential 
$\tilde{V}_c$ by utilizing completeness relations that sum over spin 
($s$) and polarization states ($r$),
\begin{align}
 \sum_s u_\mu(p,s)\bar{u}_\nu(p,s)=\Big(\frac{\not{p}+m}{2m}\Big)_{\mu\nu}\ ,
 \label{spinorsum}
 \end{align}
 \begin{align}
 \sum_r\epsilon^*_\mu(p,r)\epsilon_\nu(p,r)=g_{\mu\nu}-\frac{p_{\mu} p_{\nu}}{m^2}\ .
 \label{vecpolar}
\end{align}
The spinor terms in Eq.~(\ref{eqn:V}) are evaluated by first replacing the 
$u$'s with the completeness relation from 
Eqs.~(\ref{spinorsum},\ref{vecpolar}),
\begin{align}
\nonumber\Pi=|\bar{u}(p_3)\gamma&^\mu u(p_1)\epsilon(p_4)_{\rho}^*\Gamma^{\rho\lambda}_\mu\epsilon(p_2)_\lambda|^2 \\
=&\frac{\text{Tr}\{(\not{p_3}+m_1)\gamma^\mu(\not{p_1}+m_1)\gamma^\sigma\}}{4m_1^2}\\
&\nonumber\times\Gamma^{\rho\lambda}_\mu\Gamma^{\alpha\beta}_\sigma E_{\alpha\beta}E_{\rho\lambda}\ ,
\end{align}
where the $\Gamma$ and $E$ tensors are defined as
\begin{align}
 \nonumber\Gamma^{\rho\lambda}_\mu=&-(-g^{\rho\lambda}(p_4+p_2)_\mu+g^{\lambda}_\mu(p_2-q)^\rho\\
 &+g^{\rho}_\mu(q+p_4)^\lambda)\ ,\\
  E_{\mu\nu}=&g_{\mu\nu}-\frac{p_\mu p_\nu}{m_2^2}  \ .
 \end{align}
Contracting across the $\Gamma$'s and a term being 
traced over with computational tools~\cite{Mertig1991345}, we utilize 
standard Mandelstam variables to simplify the 
result,
\begin{align}
s=&(p_1+p_2)^2\rightarrow p_1\cdot p_2=\frac{1}{2}(s-p_1^2-p_2^2) \ ,
\label{eqn:s}
\\
t=&(p_1-p_3)^2\rightarrow p_1\cdot p_3=-\frac{1}{2}(t-p_1^2-p_3^2) \ ,
\label{eqn:t}
\\
u=&(p_1-p_4)^2\rightarrow p_1\cdot p_4=\frac{1}{2}(-s-t
\nonumber \\
&+m_1^2+m_2^2+m_3^2+m_4^2-p_1^2-p_4^2) \ ,
\label{eqn:u}
\end{align}
where for the last line we have used $s+t+u=m_1^2+m_2^2+m_3^2+m_4^2$. Furthermore, 
we employ the on-shell conditions $p^2=m^2$ for in- and outgoing particles. 
Following Ref.~\cite{Riek:2010fk}, we drop terms of order $t$ or higher  
relative to $s$, $m_Q^2$. This is justified in the high-energy elastic limit 
which is the one of interest to infer the relativistic corrections. We 
now have an expression with $\sim$200 terms that must be contracted with 
the remaining vector-boson completeness relation. An important feature
for the thermal gluon is that it should not exhibit the 3 degrees of 
freedom expected from a massive spin-1 vector-boson, since the 
longitudinal degree of freedom is not observed in the high-temperature
limit of the QGP equation of state computed in 
lQCD~\cite{PhysRevC.57.1879,Cheng:2007jq}. To enforce this 
characteristic, we project out the transverse modes with the help of 
the usual projection operators,
\begin{align}
(P_{T}+P_{L})_{\mu\nu}=g_{\mu\nu}-\frac{p_\nu p_\mu}{m^2} \ ,
\\
(P_T)_{ij}=\delta_{ij}-\frac{p_i p_j}{m^2} \ ,
\\
(P_T)_{00}=(P_T)_{0i}=(P_T)_{j0}=0 \ ,
\end{align}
where $i=1,2,3$. By selecting $P_T$, we proceed with the contraction 
that schematically is viewed as
\begin{align}
\Pi_T=\Lambda^{ijkl}(P_T)_{ij}(P_T)_{kl}=\Lambda^{ijkl}E_{ij}E_{kl} \ ,
\end{align}
with $\Lambda$ representing the result of the first contraction. We are 
then left with a series of purely 3-D scalar products.  To see this, we 
rework the Mandelstam relations for 3-momenta,
\begin{align}
\mathbf{p}_1\cdot \mathbf{p}_2=\frac{1}{2}(-s+m_1^2+m_2^2+2\omega_1\omega_2) \ ,
\\
\mathbf{p}_1\cdot \mathbf{p}_3=-\frac{1}{2}(t-m_1^2-m_3^2+2\omega_1\omega_3) \ ,
\\
\mathbf{p}_1\cdot \mathbf{p}_4=\frac{1}{2}(-s-t+m_2^2+m_3^2+2\omega_1\omega_4) \ ,
\end{align}
where $E_i=\omega_i=\sqrt{m_i^2+\mathbf{k}_i^2}$. In the center of mass (CM)
of an elastic collision, we set $\omega_{1/2}=\omega_{3/4}$ and 
$\mathbf{p}_{1/3}=-\mathbf{p}_{2/4}$.
Thus,
\begin{align}
\nonumber&\Pi_T=\\
\nonumber &\frac{2}{m_1^2} \Big((-{\mathbf{p}_1}^2 ({m_1}^2-{\omega_1}^2) ({m_2}^2+{m_1}^2-s+2 {\omega_1} {\omega_2})^2\\
\nonumber&+({m_2}^2-{\omega_2}^2)^2 ({m_2}^2+{m_1}^2-s)^2+2 {\mathbf{p}_1}^8) \frac{1}{{\mathbf{p}_1}^4}\\
&-4 {\omega_1} {\omega_2} ({m_2}^2+{m_1}^2-s)-2 ({m_1}^2-{\omega_1}^2)^2-4 {\omega_1}^2 {\omega_2}^2\Big)\ ,
\end{align}
which, upon averaging over spin states, can be further simplified to 
\begin{equation}
\Pi_T=\frac{1}{m_1^2}(s-m_2^2-m_1^2)^2  \ .
\end{equation}
We take $\Pi_T$ as the result of the spinor structure calculation 
and ignore the longitudinal mode. We set $m_{2/4}=m_g$ and $m_{1/3}=m_Q$. 
Referring back to Eq.~(\ref{eqn:Vrel}), our modified potential, 
$\tilde{V}_c$, thus reads
\begin{align}
\tilde{V}_c=V_c\sqrt{\frac{(s-m_g^2-m_Q^2)^2}{4m_g^2m_Q^2}} \ .
\end{align}
We see that $\Pi_T/4m_g^2$ yields the same corrections as found for
the heavy-light quark case elaborated in Ref.~\cite{Riek:2010fk}; we adopt 
the notation in there to summarize our relativistic corrections as 
\begin{align}
R(q,q')=&m(q)^{-1/2}m(q')^{-1/2}\label{eqn:R} \ ,
\\
B(q,q')=&b(q)^{1/2}b(q')^{1/2} \ ,
\\
b(q)=&(1+\frac{q^2}{\omega_g(q)\omega_Q(q)}) \ ,
\\
m(q)=&\frac{m_Qm_g}{\omega_Q(q)\omega_g(q)} \ .
\label{mq}
\end{align}
The relativistically augmented Coulomb potential for HQ-gluon
scattering then reads
\begin{align}
\tilde{V}_c=V_c \ R(q,q') \ B(q,q') \ .
\end{align}

We now perform the same analysis on the string portion of the potential
which requires an ansatz for a pertinent Lagrangian. Assuming the string
interaction to be of scalar type one has 
\begin{equation}
\mathcal{L}=\tilde{G}m_gA^\mu A_\mu \bar{Q}Q \ ,
\label{Lstring}
\end{equation}
which is characterized by a dimensionful coupling, $\tilde{G}$, which in 
momentum space recovers the Fourier transform of the static linear 
coordinate-space potential, $V_s\sim m_G^2/(\mathbf{k}^2+\tilde{m}_D^2)^2$. 
As in the Coulomb case, we can then write down the amplitude (in 
Bjorken-Drell normalization) as a potential with relativistic
corrections from the spinor terms, 
\begin{align}
\tilde{V}_s=\frac{V_s}{2m_g}
|\bar{u}(p_3)u(p_1)2m_g \epsilon(p_4)_{\mu}^*\epsilon(p_2)^\mu| \ .
\label{stringcorrections}
\end{align}
Calculating the spinor terms with transverse gluons yields
\begin{align}
\nonumber\Xi=&|\bar{u}(p_3)u(p_1)2m_g g^{\mu\nu}\epsilon(p_4)_{\mu}^*\epsilon(p_2)_\nu|^2\\
\nonumber\Xi_T=&\text{Tr}\{(\not{p_1}+m_1)(\not{p_3}+m_3)\}E_{ij}E^{ij}\frac{4m_g^2}{4m_Q^2}\\
=&(8m_Q^2-2t)(1+\frac{(\bold{p}_2\cdot\bold{p}_4)^2}{\bold{p}_2^2\bold{p}_4^2})
\frac{4m_g^2}{4m_Q^2} \ .
\label{eqn:xi}
\end{align}
We can simplify Eq.~(\ref{eqn:t}) by plugging our elastic condition, 
$t=-(\mathbf{p}_2-\mathbf{p}_4)^2$, into Eq.~(\ref{eqn:xi}),
\begin{equation}
\Xi_T=8m_g^2(1+\frac{1}{4}(\bold{p}_2^4+\bold{p}_4^4+2\bold{p}_2^2\bold{p}_4^2)
(\bold{p}_2^2\bold{p}_4^2)^{-1}) \ .
\label{eqn:xi2}
\end{equation}
Utilizing CM kinematics, $\bold{p}_2^2=\bold{p}_4^2$, Eq.~(\ref{eqn:xi2}) 
reduces to $\Xi_T=16m_g^2$.  Averaging over spin states and dividing through 
by the normalization factor, one obtains
\begin{equation}
\tilde{V}_s=V_s\frac{\sqrt{\Xi_T/4}}{2m_g}=V_s \ .
\end{equation}
The net result is that the string potential does not develop a relativistic 
correction within our approximation scheme (again, paralleling the heavy-light
quark case~\cite{Riek:2010fk}). 

The parameterization of the static lQCD coordinate potential in principle 
includes the effect of a running coupling constant at short distances.
However, its off-shell extension in the Fourier transform does not, i.e.,
if the in- and outgoing moduli of the relative 3-momenta are different. We
therefore introduce a correction to simulate the off-shell running of 
$\alpha_s$ through a factor
\begin{equation}
\text{F}_{\rm run}(q,q')=\text{ln}
[\frac{\Delta^2}{\Lambda^2}]/\text{ln}[\frac{(q-q')^2+\Delta^2}{\Lambda^2}] \ ,
\end{equation}
with $\Delta=1$\,GeV and $\Lambda=0.2$\,GeV.  

We are now in position to quote the final form  our relativistically 
generalized static potential as
\begin{equation}
\tilde{V}(q,q')=R(q,q')B(q,q')F_{\rm run}(q,q')V_c(q,q')+V_s(q,q') \ .
\end{equation}
Its explicit decomposition into color projections and partial waves  
reads
\begin{align}
\nonumber\tilde{V}^{l,a}(q,q')=R(q,q')B(q,q')F_{\text{run}}(q,q')V_c^{l,a}(q,q')\\
+V_s^l(q,q')\ ,
\label{Vla}
\end{align}
where $l=0,1,2...$ denotes the angular-momentum quantum number and $a$ 
the color index of the irreducible representations of the triplet-octet 
system (the Coulomb part follows Casimir scaling while the string term 
is assumed to be color-blind). As in previous work we neglect spin-orbit 
and and spin-spin contributions to the potential.  
Equation (\ref{Vla}) is the explicit expression to be used in the 
pertinent $T$-matrix equations for each combination of $\{l,a\}$.  

\section{$T$-Matrix}
\label{sec_tmat}
We solve for the $\tilde{T}$-matrix utilizing a 3-D reduced Bethe-Salpeter \cite{PhysRev.84.1232} 
equation in ladder approximation \cite{Machleidt:1989tm,Thompson:1970wt}. After partial-wave expansion and 
color projection we obtain a 1-D integral equation which yields itself 
to numerical analysis, 
\begin{eqnarray}
\tilde{T}^{l,a}(E;q',q)=\tilde{V}^{l,a}(q',q)+\frac{2}{\pi}
               \int\limits_0^\infty dkk^2\tilde{V}^{l,a}(q',k)
\nonumber\\
\times G_{12}(E;k)\tilde{T}^{l,a}(E;k,q)[1- n_1(\omega_1(k)) + 
n_2(\omega_2(k))] \ . 
\label{tmatrix}
\end{eqnarray}
The thermal distribution functions, $n_{i}$, figure according to their
quantum statistics, i.e., fermion for $i$=1 (heavy quark) and boson
for $i$=2 (gluon). At the relevant QGP temperatures of up to 2-3\,$T_c$
their impact is small (negligible for heavy quarks). As before, 
$q=|\mathbf{q}|$ and $q'=|\mathbf{q}'|$ denote the relative 3-momenta of 
the incoming and outgoing states, respectively, and $k=|\mathbf{k}|$ is 
the relative (integration-) momentum of the intermediate states.
To complete the analysis of our $\tilde{T}$-matrix, we must select a 
propagator which amounts to specifying the reduction scheme of the underlying
4-D scattering equation. For the present paper we employ the Thompson \cite{Thompson:1970wt} scheme,
\begin{align}
G_{gQ}(q)=&\frac{m(q)}{E-\omega_g(q)-\omega_Q(q)-\Sigma_g-\Sigma_Q} \ ,
\end{align}
where $\omega_{g,Q}(q)=\sqrt{q^2+m_{g,Q}^2}$, and $m(q)$ is defined in 
Eq.~(\ref{mq}). We motivate the choice of 
the Thompson scheme by stating that it has been previously 
suggested~\cite{Woloshyn:1973} to provide the closest resemblance to the
original Bethe-Salpeter equation. As in Ref.~\cite{Riek:2010fk} the masses 
of the heavy quarks are determined by the sum of a bare mass, $m_Q^0$, 
and an in-medium contribution defined by the infinite-distance limit of 
the potential (free or internal energy), 
\begin{equation}
m_Q=m_Q^0+\Sigma_Q^R(T)\hspace{.06in},\hspace{.06in} 
\Sigma_Q^R(T) = X(T,\infty)/2 \ ,
\end{equation}
with $X=F \text{ or } U$. The bare mass is adjusted to the ground-state
quarkonium mass in vacuum (where the self-energy contribution, 
$\Sigma_Q^R(T=0)$, is evaluated at a typical string-breaking scale of
$r\simeq$1-1.2\,fm). When using the lQCD potential of 
Ref.~\cite{Kaczmarek:2007pb} within the Thompson scheme one finds
$m_c^0=1.264$\,GeV  and $m_b^0=4.662$\,GeV. 
 The in-medium gluon mass, $m_g$, is approximated by 
its expression from thermal perturbation theory~\cite{bellac2000thermal},
\begin{equation}
m_g^2 =\frac{g^2T^2}{2}(\frac{N_c}{3}+\frac{N_f}{6}) \ .
\end{equation}
We choose 3 active light flavors ($N_f=3$), and with three colors, 
$N_c=3$, the thermal mass correction is $m_g=\sqrt{{3}/{4}}gT$. We fix
$g=2.3$, which is consistent with lQCD calculations as outlined 
in Ref.~\cite{Riek:2010fk,Cheng:2010}.
In the QGP, both heavy quarks and light partons are expected to 
acquire a substantial width. For charm quarks, it has been computed 
selfconsistently in Ref.~\cite{Riek:2010py}. For simplicity, we fix the 
combined width of heavy quark and gluon at $0.2$\,GeV, i.e., $0.1$\,GeV
for each particle. This approximation has been shown previously to give
good agreement with the selfconsistent results when computing the HQ
relaxation rates, which is our main focus here.

\begin{figure*}
\includegraphics{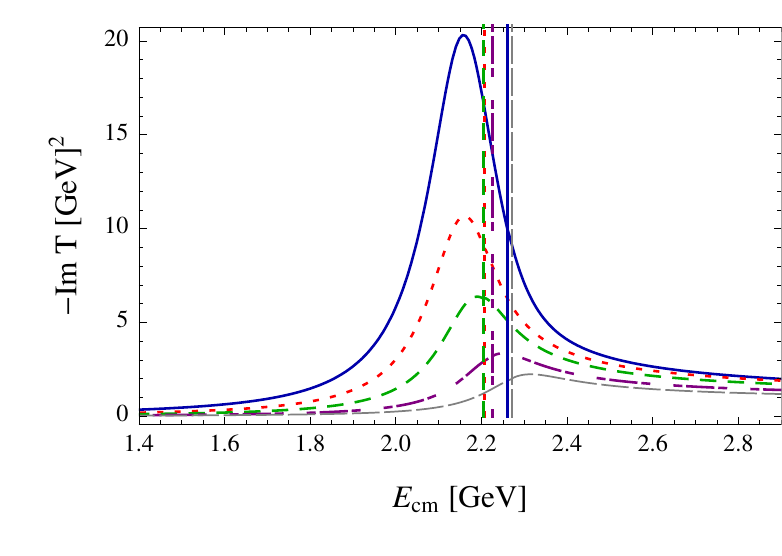}
\includegraphics{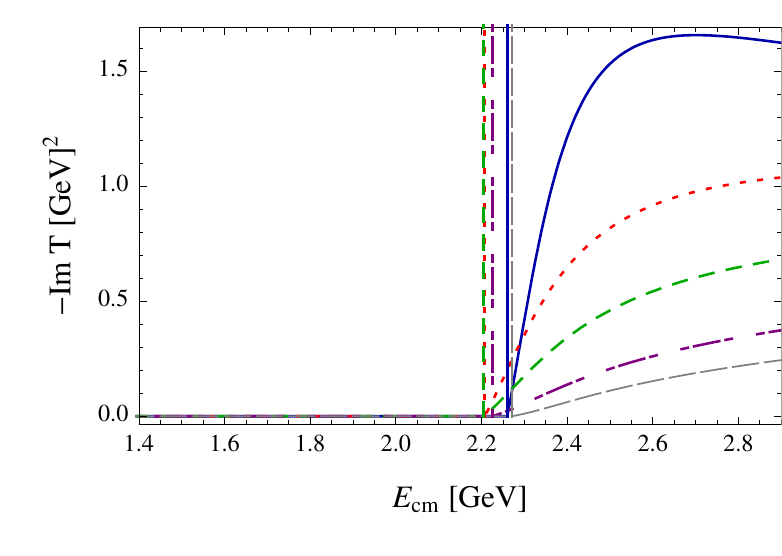}\\
\includegraphics{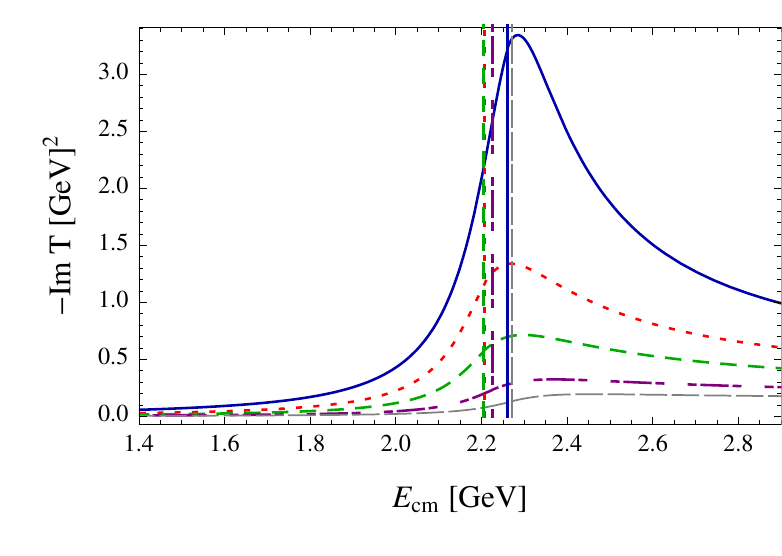}
\includegraphics{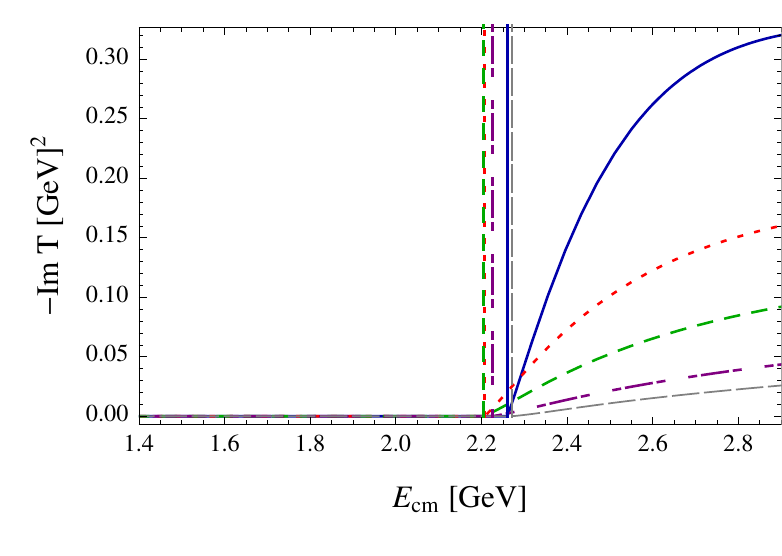}\\
\includegraphics{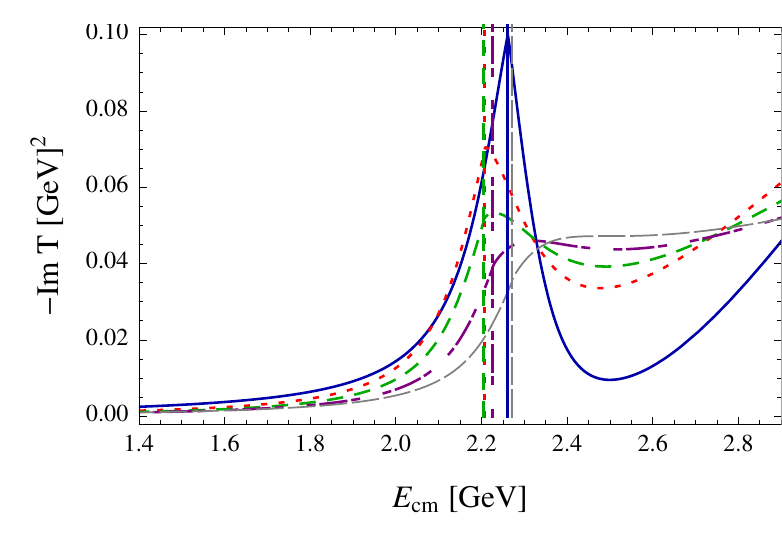}
\includegraphics{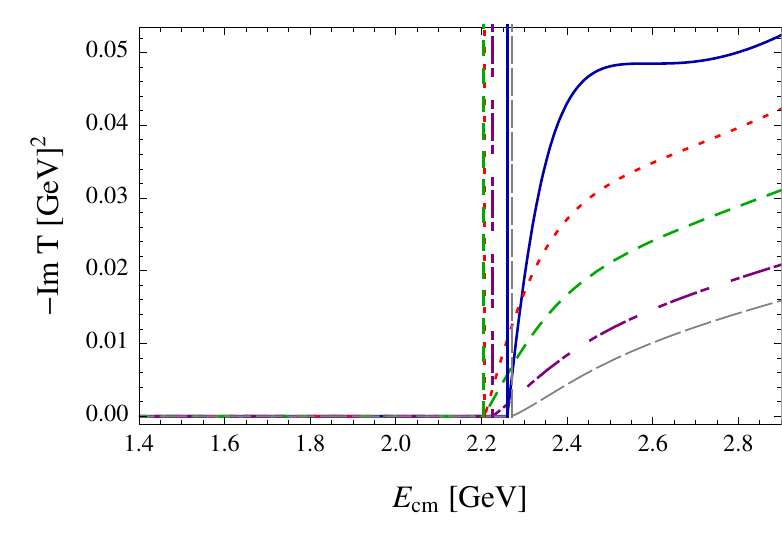}
\includegraphics[scale=.5]{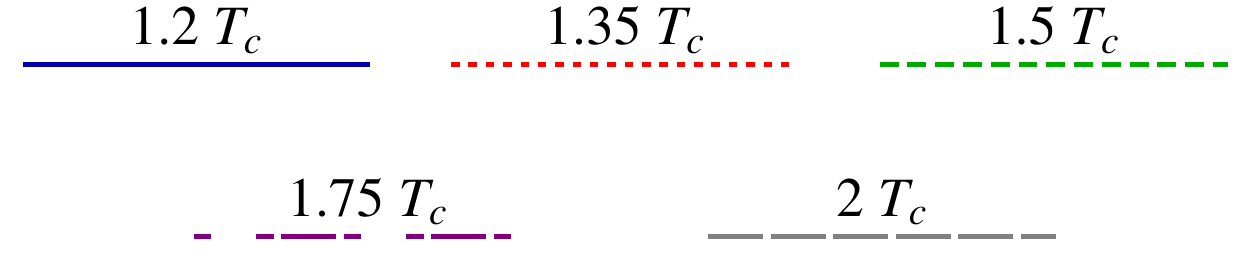}
\caption{Imaginary part of the on-shell charm-gluon $\tilde{T}$-matrix for 
$S$-wave (left) and $P$-wave (right) scattering in the triplet (upper), 
sextet (middle), and 15-plet (bottom) channels using $U$ as a potential and 
a fixed two-particle width of 200\,MeV. Vertical lines indicate the mass threshold
at the associated temperature. (Color online.)}
\label{TU}
\end{figure*}
\begin{figure*}
\includegraphics{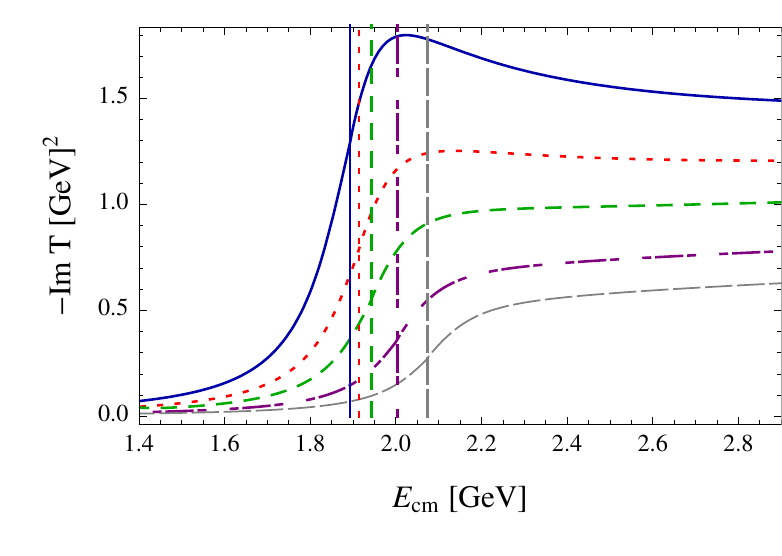}
\includegraphics{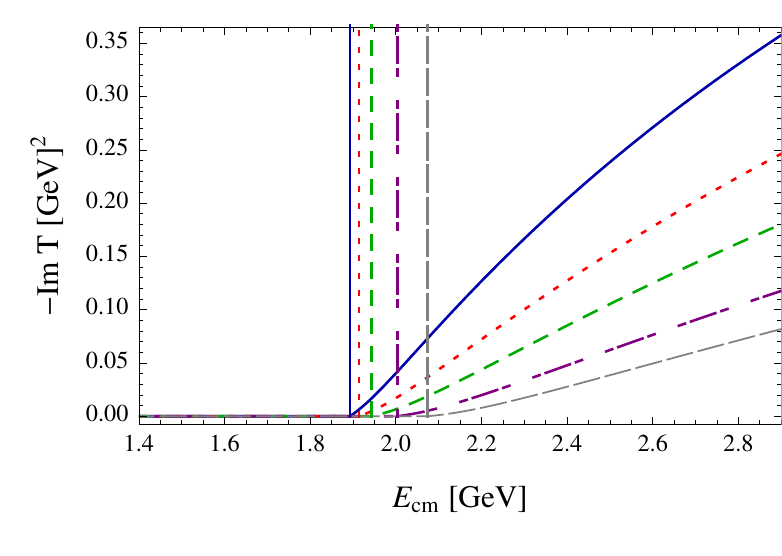}\\
\includegraphics{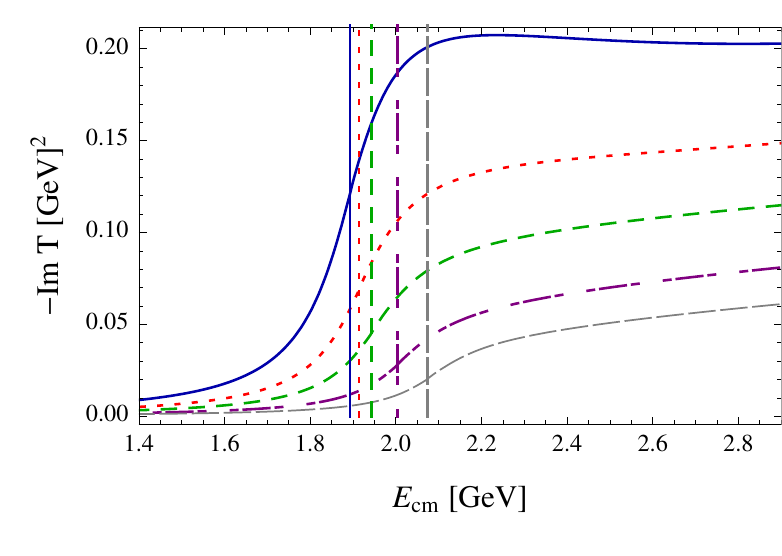}
\includegraphics{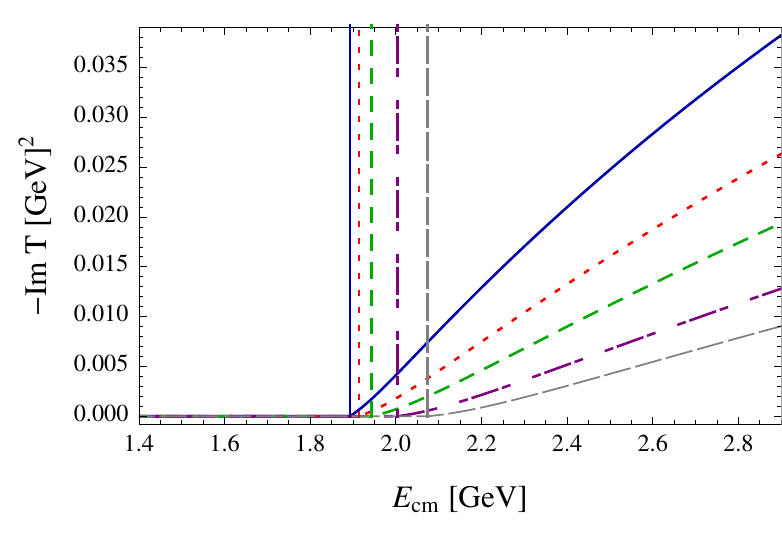}\\
\includegraphics{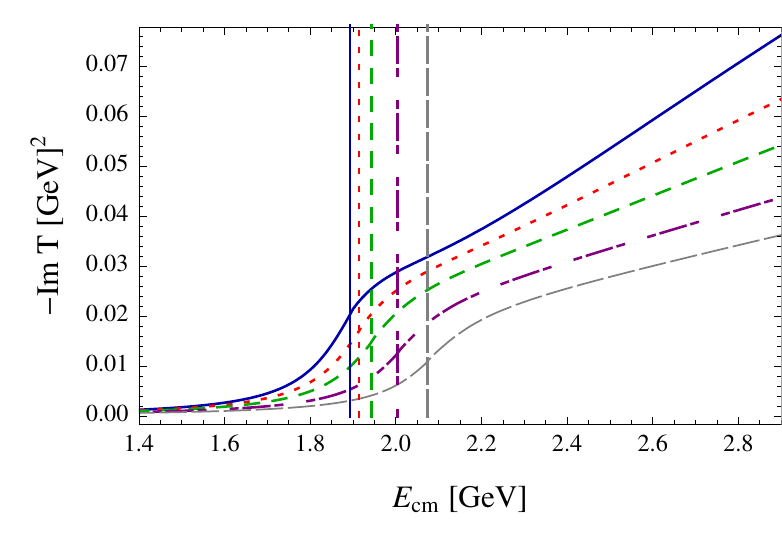}
\includegraphics{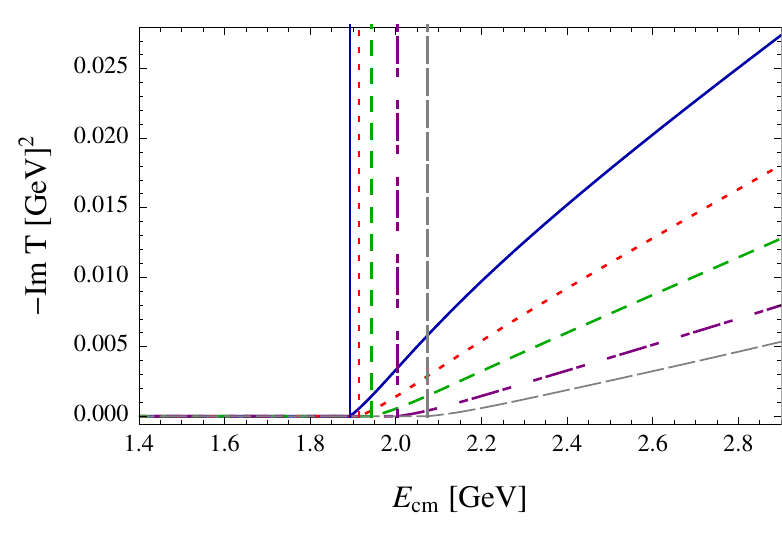}
\includegraphics[scale=.5]{images/TmatrixPlots/TmatrixLegend.pdf}
\caption{Same as Fig.~\ref{TU}, but using potential $F$.}
\label{TF}
\end{figure*}
\begin{figure*}[t]
\includegraphics{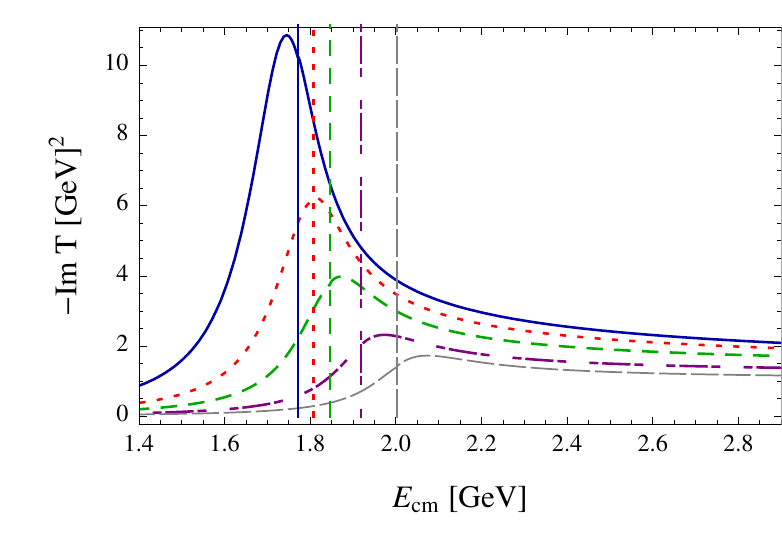}
\includegraphics{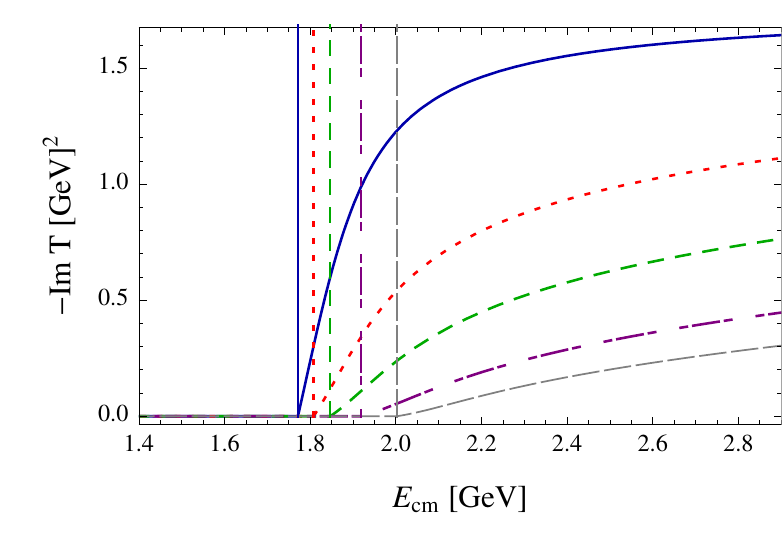}\\
\includegraphics{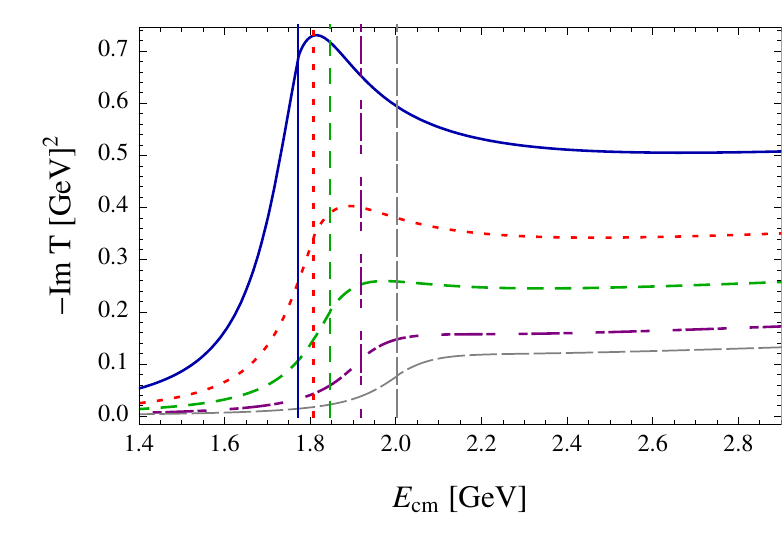}
\includegraphics{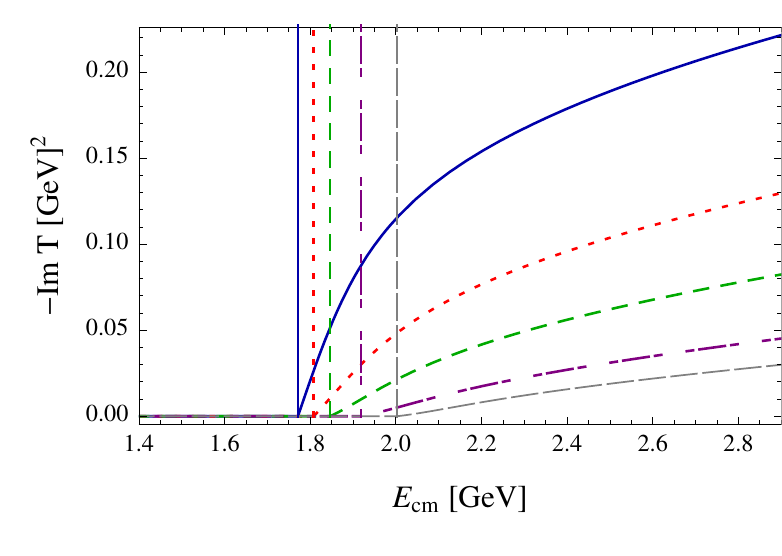}
\includegraphics{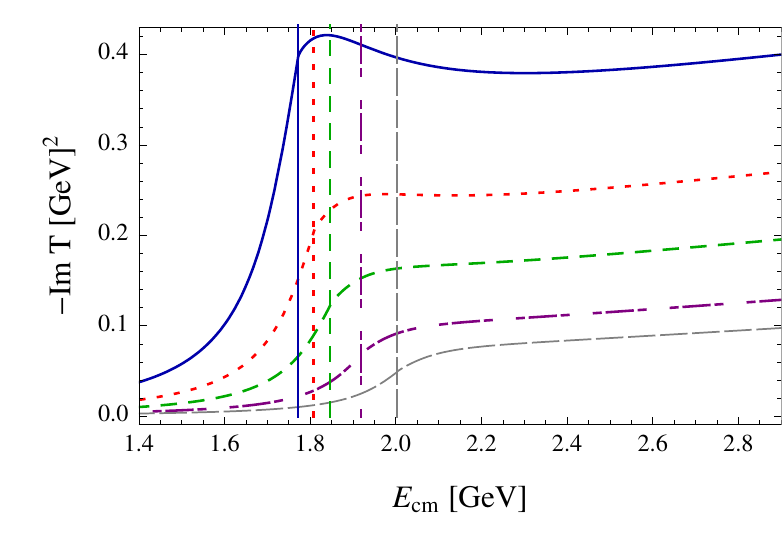}
\includegraphics{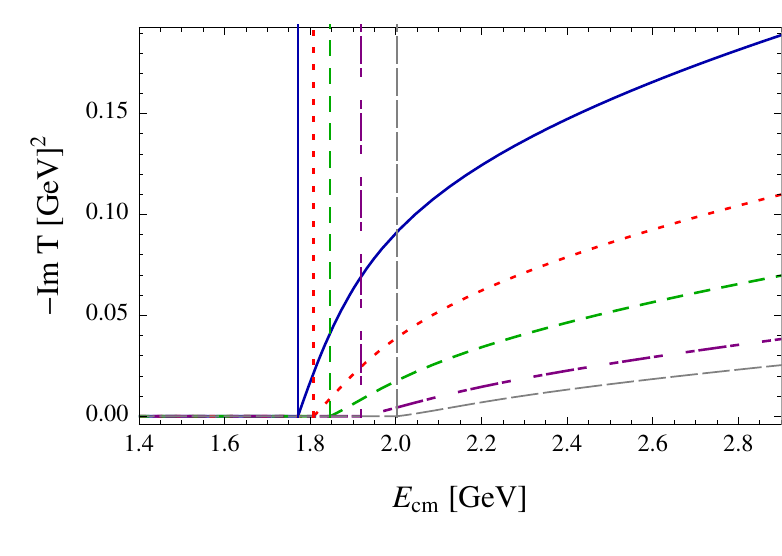}
\includegraphics[scale=.5]{images/TmatrixPlots/TmatrixLegend.pdf}
\caption{Same as Fig.~\ref{TU}, but for a potential with only the
Coulomb term.}
\label{CcT}
\end{figure*}

The resulting $\tilde{T}$-matrices, depicted in Fig.~\ref{TU}, indicate 
near-threshold resonances in the attractive triplet and sextet channels 
up to temperatures of $\sim$1.5\,$T_c$. The repulsive 15-plet channel is
suppressed in strength by more than an order of magnitude in the near-threshold
regime, and turns out to be comparable in $S$- and $P$-waves. On the other 
hand, in the attractive color channels the $P$-waves are substantially 
suppressed relative to the $S$-waves not too far above threshold, but 
become comparable well above threshold once the $S$-wave resonance 
structure ceases. When changing the input potential to the free energy, 
$F$, (cf.~Fig.~\ref{TF}), the 2-particle threshold energy is significantly 
reduced at low temperatures. Due to the reduced interaction strength, the 
attractive resonance structures universally disappear in all channels for 
all temperatures (excluding perhaps the triplet at $1.2T_c$); the
remaining enhancements in the $S$-waves now mostly occur above the 
threshold energies. The latter continually increase with temperature, 
as opposed to the non-monotonic behavior for the internal energy. This 
is attributed to the dominance of the increasing thermal-mass contribution 
of the gluon over the weakly decreasing infinite-distance limit of the 
free energy governing the in-medium HQ mass. 

To investigate the relative importance and interplay of the Coulomb and
string parts in our potential ansatz, we perform identical calculations 
for the Coulomb portion only by setting $m_G=0$. This, in particular, 
implies that the infinite-distance limit of the potential is largely 
reduced, leading to a smaller mass correction (which can even become 
negative as expected from thermal perturbation theory at high $T$).
The smaller mass enhances the potential through the relativistic 
correction factors in Eq.~(\ref{eqn:R}) entailing an increase in the 
$\tilde{T}$-matrix. On the other hand, the attractive channels experience
a loss of attraction due to the missing string term, while in the 
repulsive channel the compensation effect with the attractive string
term is lifted. These effects are quantitatively borne out of 
Fig.~\ref{CcT}. The Coulomb-only $\tilde{T}$-matrices are reduced by up 
to a factor of 2 (4) in the triplet (sextet) channels in the resonance 
region, and less so at higher energies and with increasing $T$. On the 
other hand, the 15-plet increases 
by a factor of up to $\sim$4 close to threshold and close to $T_c$. 
The sextet and 15-plet thus exhibit the strongest sensitivity since
here Coulomb and string term are comparable and thus their interplay is
most pronounced (enhancement and compensation, respectively). 
Another important component is the threshold energy: in the full potential 
at 1.2\,$T_c$ the charm-quark mass is $\sim$1.8\,GeV, while for Coulomb 
only it is $\sim$1.3\,GeV. This increases the phase space allowing for 
more possible interactions when the particle propagates in the medium 
which has important consequences for the HQ transport coefficient.  
Additionally, in the attractive channels, the Coulomb $\tilde{T}$-matrix 
peaks somewhat closer to threshold (due to slightly less binding), 
thus making available more interaction strength in the scattering regime
(continuum). In the $P$-wave, the differences in magnitude of the
$\tilde{T}$-matrix between the full and Coulomb-only calculations are 
less pronounced in the attractive channels (higher momenta are probed 
where the string term is suppressed), while the 15-plet still shows a 
noticeable enhancement.
These dynamic and kinematic effects will be important in the
interpretation of the pertinent results for the transport coefficient 
discussed below.  

\section{Transport Coefficient}
\label{sec_trans}
To examine the diffusion properties of a single heavy (anti-) quark in 
the context of our heavy-light $\tilde{T}$-matrix, we adopt the usual
Fokker-Planck approach~\cite{PhysRevD.37.2484}. The relaxation rate,
\begin{equation}
\gamma_Q=1/\tau_Q\equiv\lim_{p\rightarrow0}A(\mathbf{p})\ ,
\end{equation}
then follows from a friction coefficient given by
\begin{align}
\nonumber A(\mathbf{p})=&\frac{1}{16(2\pi)^9\omega_Q(p)}\int\frac{d^3q}{\omega_g(q)}n_F(\omega_g(q))\int\frac{d^3q'}{\omega_g(q')}\\
\nonumber&\times\int\frac{d^3p'}{\omega_Q(p')}\frac{(2\pi)^4}{d_c}\sum|\mathcal{M}|^2\delta^{(4)}(q+p-q'-p')\\
&\times(1-\frac{\mathbf{p}\cdot\mathbf{p'}}{\mathbf{p}^2}) \ .
\label{friction}
\end{align}
The initial-state averaged and final-state summed scattering amplitude is 
related to our partial-wave expanded $T$-matrix via~\cite{vanHees:2007me}
 \begin{align}
\nonumber\sum|\mathcal{M}|^2=\frac{64\pi}{s^2}(s-m_g^2+m_Q^2)^2
(s-m_Q^2+m_g^2)^2\\
\times N_f\sum_ad_a(|T^{0,a}(s)|^2+3|T^{1,a}(s)\cos{\theta_{\rm cm}}|^2)\ ,
\end{align}
where $T$ is related to $\tilde{T}$ via Eq.~(\ref{Ttilde}),
\begin{align}
T^{i,a}(s)=&m(p_{\rm cm})^{1/2}\tilde{T}^{i,a}(E;p_{\rm cm},p_{\rm cm})
m(p_{\rm cm})^{1/2}\ ,
\\
 E=&\sqrt{s}=\omega_g(p_{\rm cm})+\omega_Q(p_{\rm cm})\ ,
\\
p_{\rm cm}=&\frac{1}{2E}\sqrt{m_Q^4+(m_g^2-s)^2-2m_Q^2(m_g^2+s)} \ ,
\end{align}
with color degeneracies for HQ scattering off 
\begin{align}
\nonumber\text{anti-/quarks: }d_0=&1\ ,\hspace{.1in}d_{\bar{3}}=3\ ,\hspace{.1in}d_6=6\ ,\hspace{.1in}d_8=8\ ,\\
\text{gluons: }d_{3}=&3\ ,\hspace{.1in}d_{6}=6\ ,\hspace{.1in}d_{15}=15 \ .
\label{lab}
\end{align}
The friction coefficient in Eq.~(\ref{friction}) is now applied for 
the gluon sector. For comparison, we also recover the results for the 
light- and strange-quark sectors (where the light sector is doubly 
degenerate). 

\begin{figure*}[htp]
\includegraphics[scale=.47]{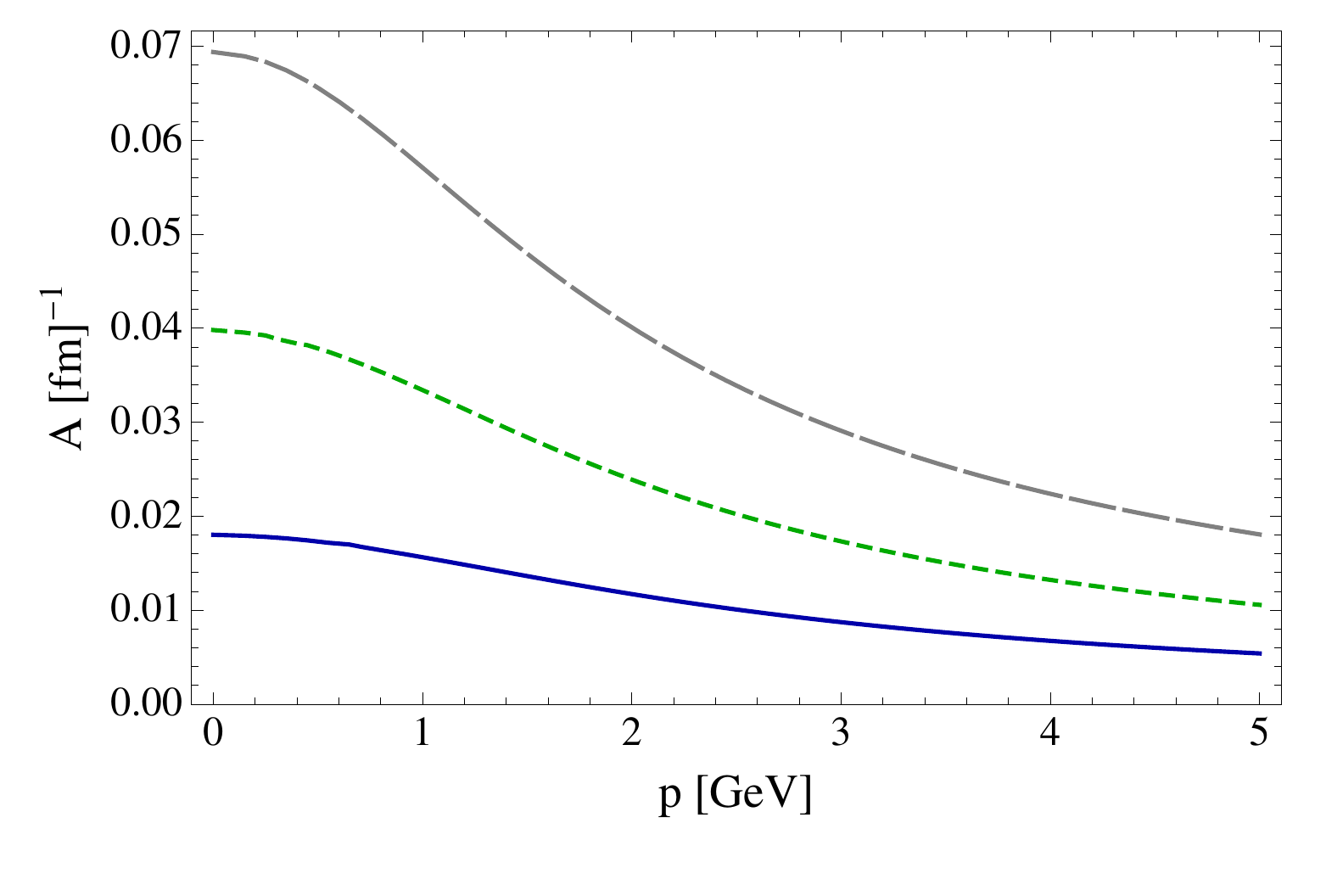}
\includegraphics[scale=.47]{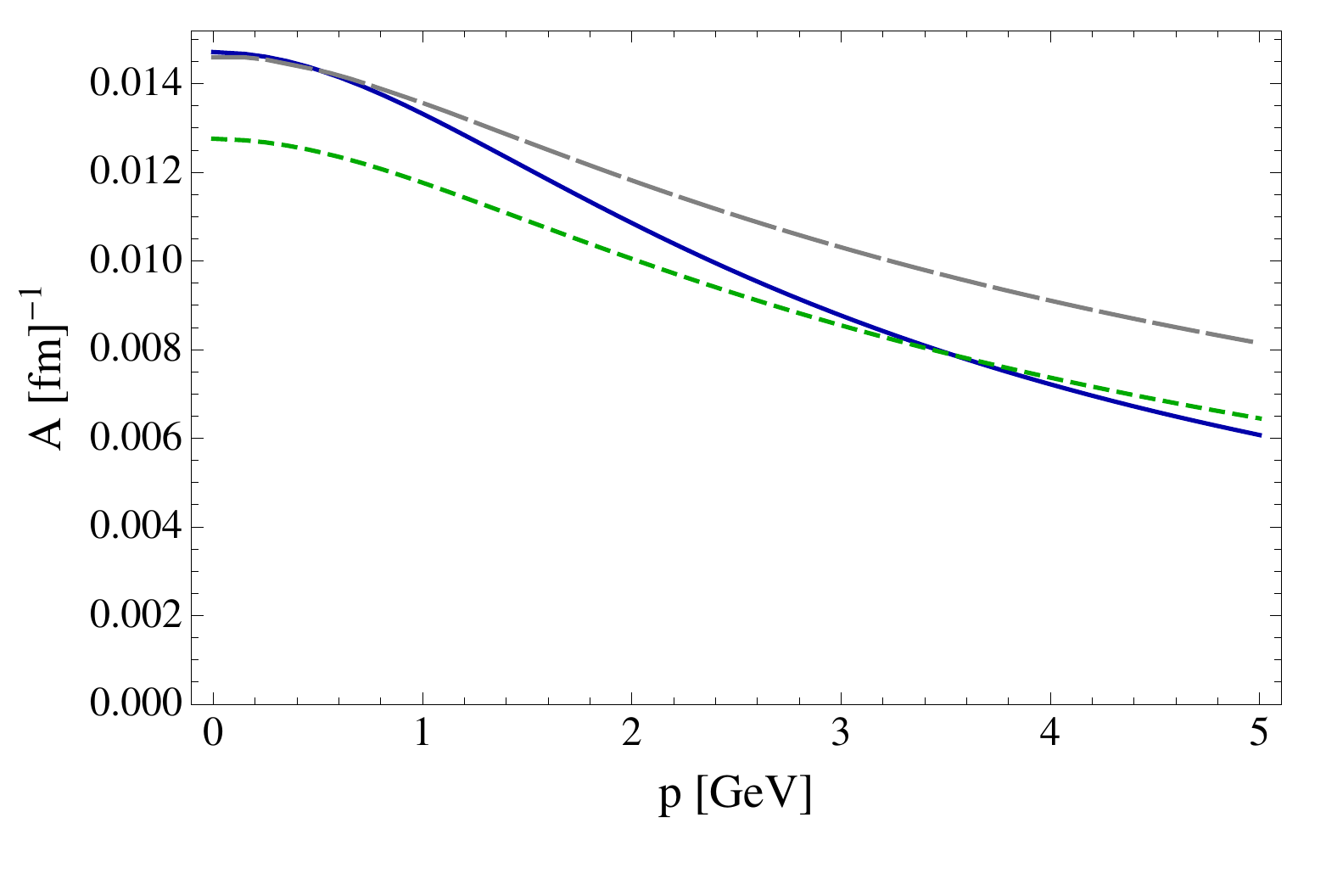}\\
\includegraphics[scale=.47]{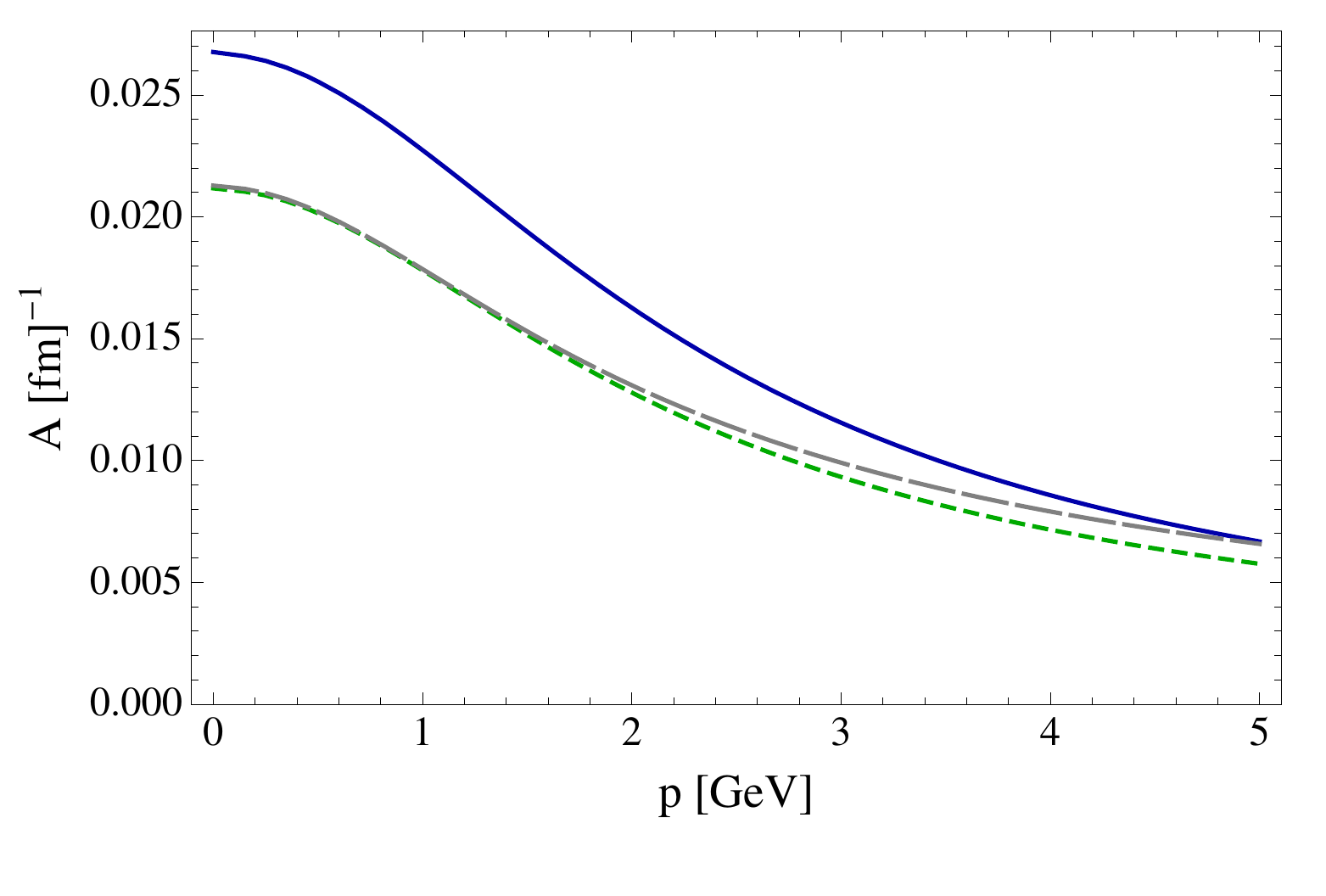}
\includegraphics[scale=.47]{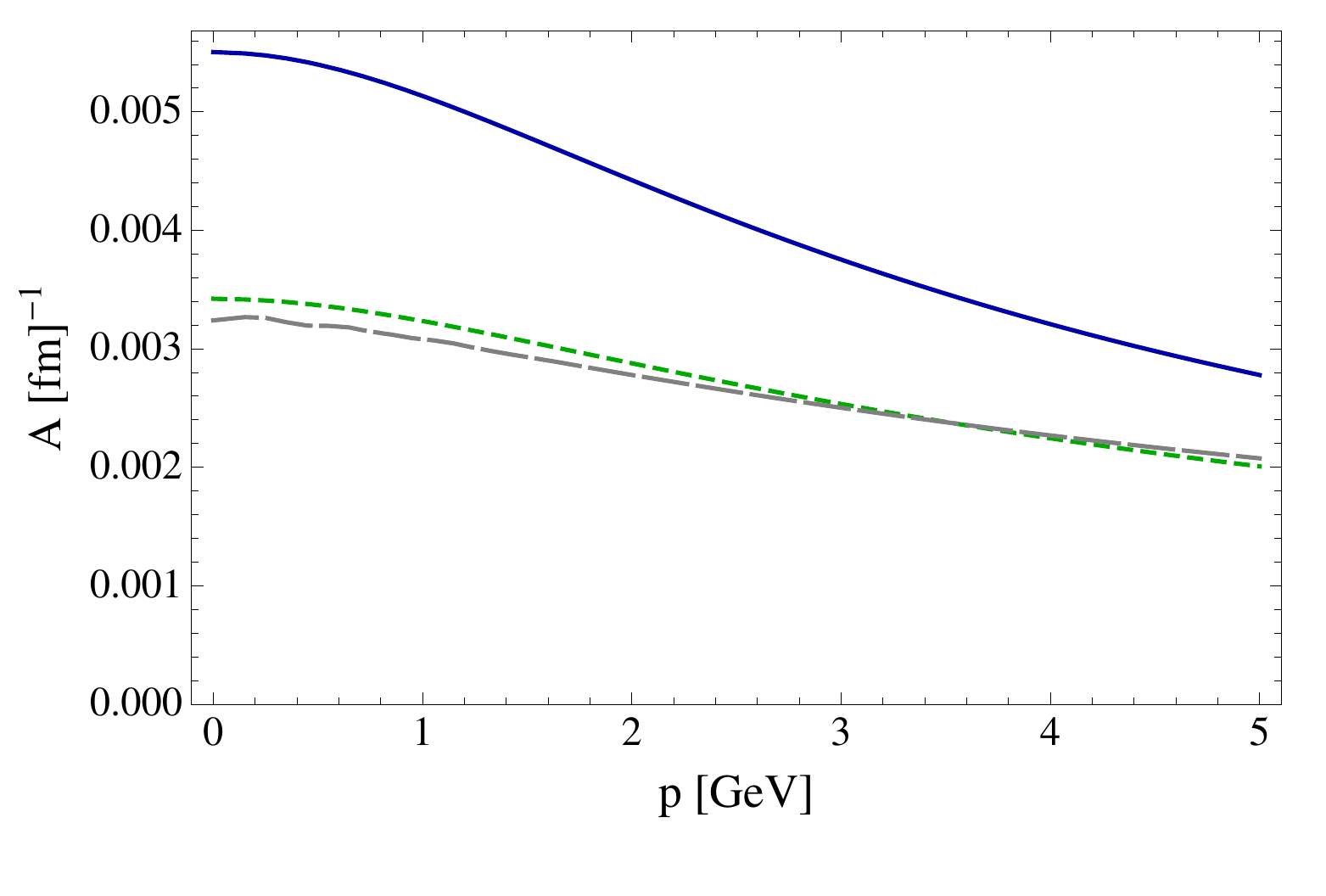}\\
\includegraphics[scale=.47]{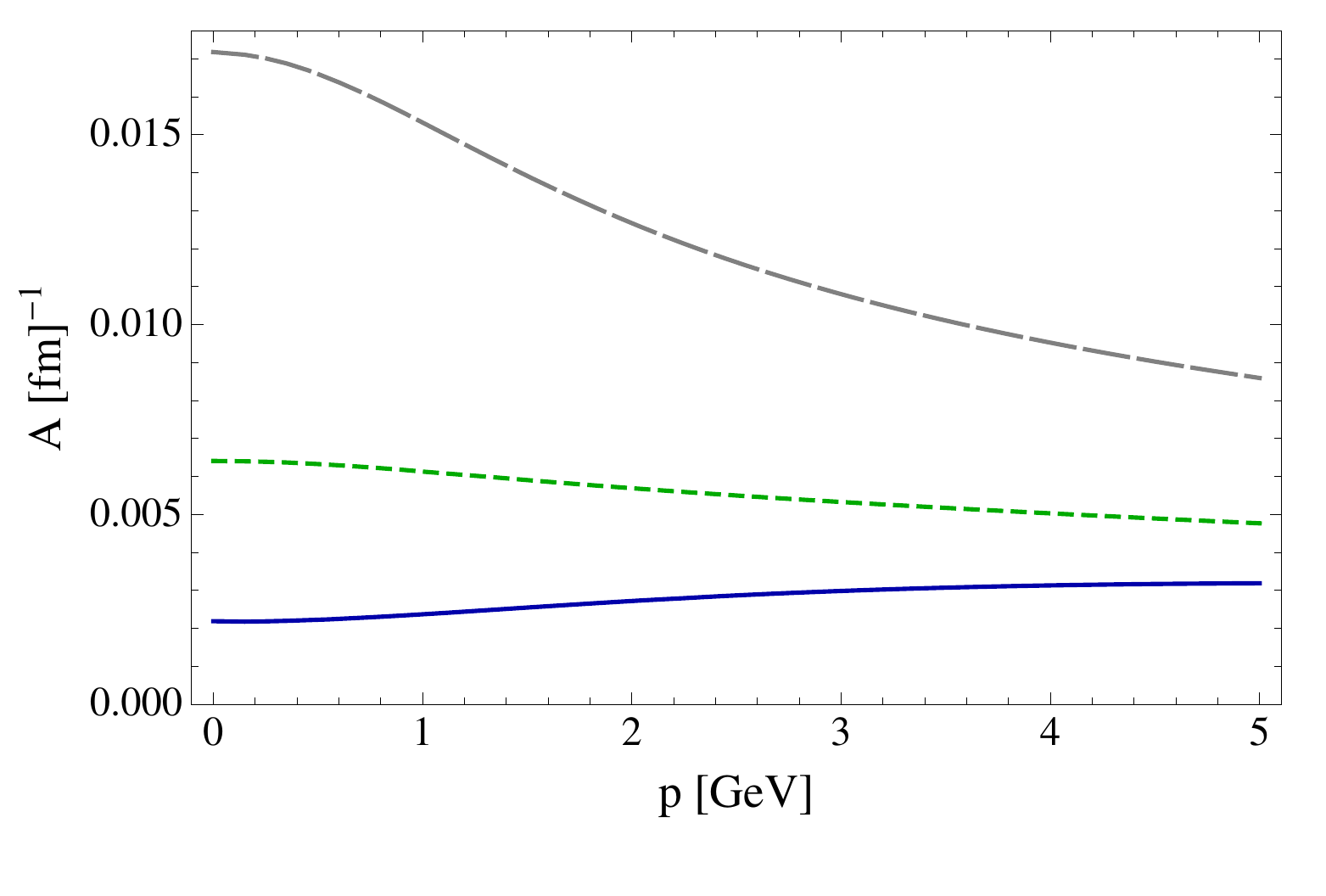}
\includegraphics[scale=.47]{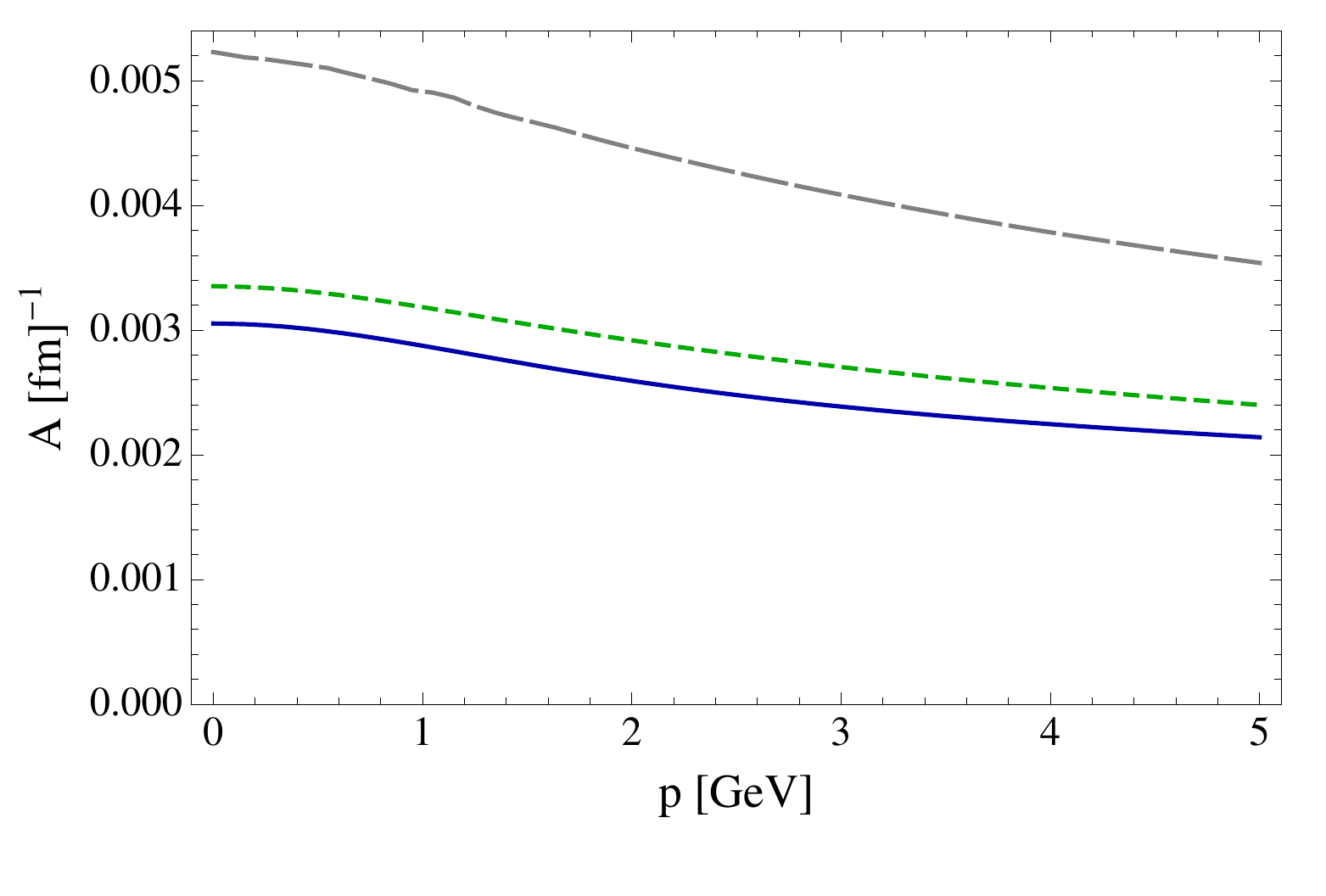}\\
\includegraphics[scale=.47]{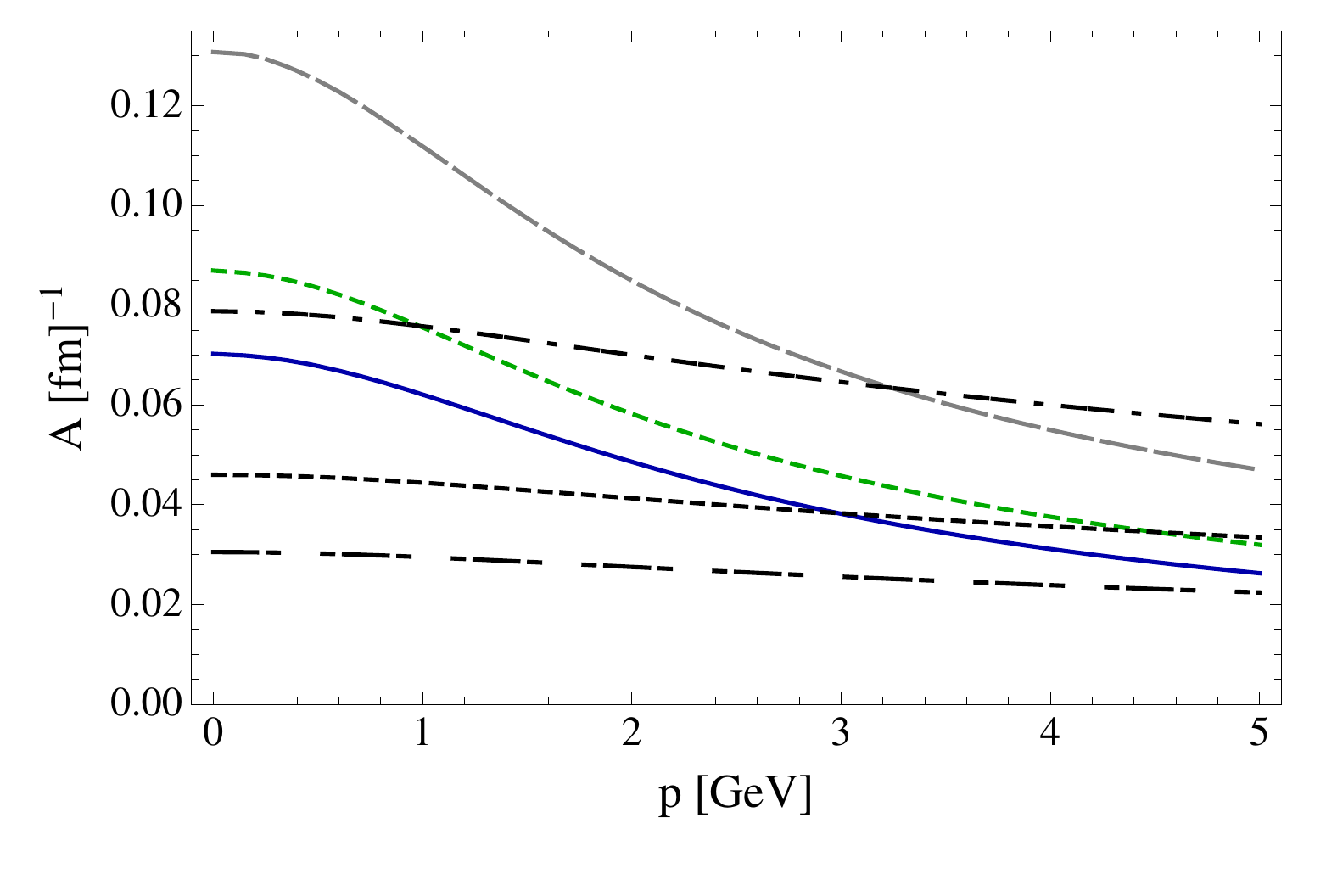}
\includegraphics[scale=.47]{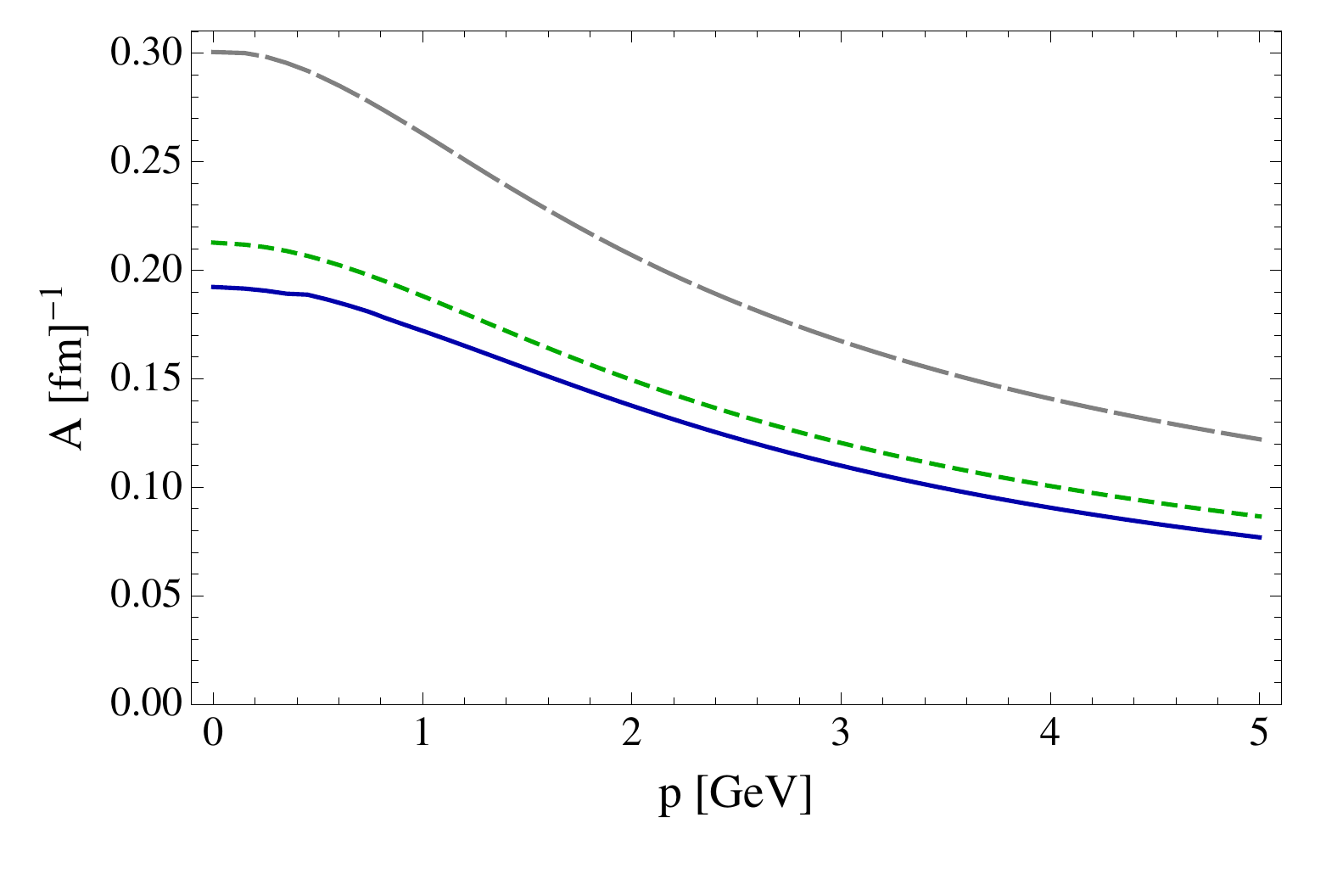}
\includegraphics[scale=.47]{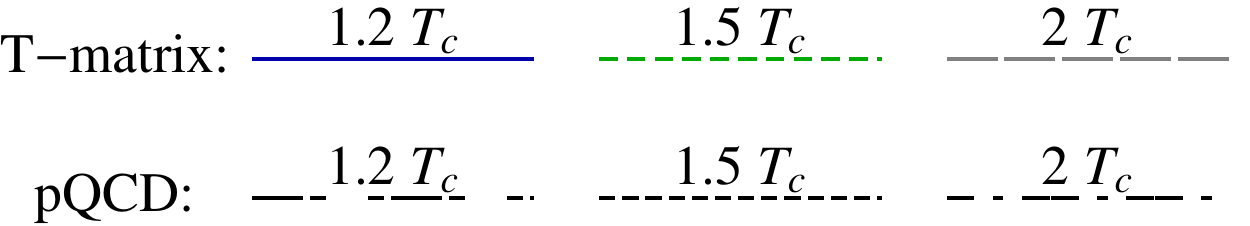}
\caption{Friction coefficients in the $S$-wave (left) and $P$-wave (right) 
for charm-gluon scattering with potential $U$ in triplet (top), sextet 
(2nd row) and 15-plet (3rd row) color channels. The sum of all gluon
contributions is displayed at the bottom left while the total sum of 
gluons, light and strange quarks is at the bottom right. (Color online.) }
\label{fig_gamc-u}
\end{figure*}

\begin{figure*}[htp]
\includegraphics[scale=.48]{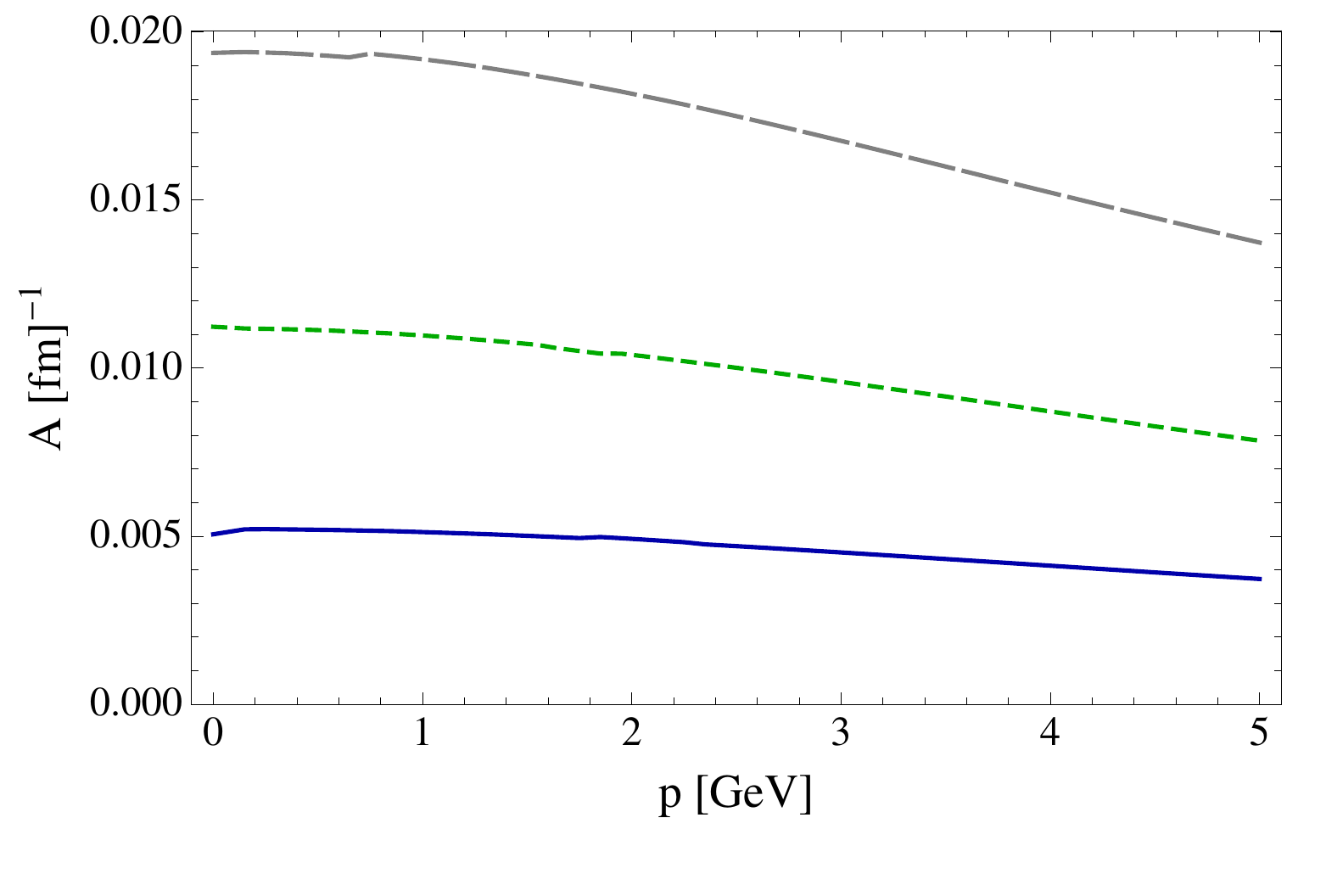}
\includegraphics[scale=.48]{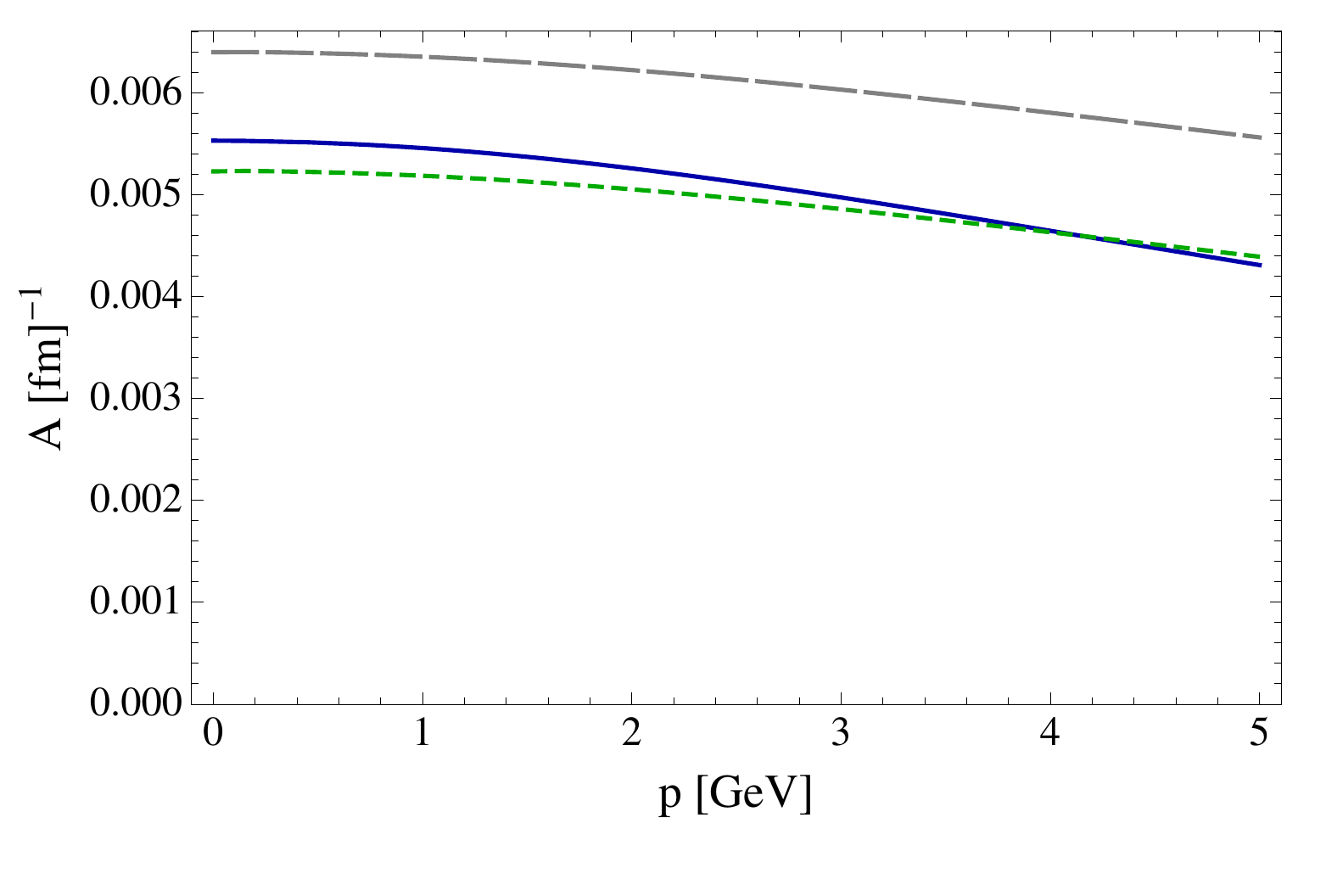}
\\
\includegraphics[scale=.48]{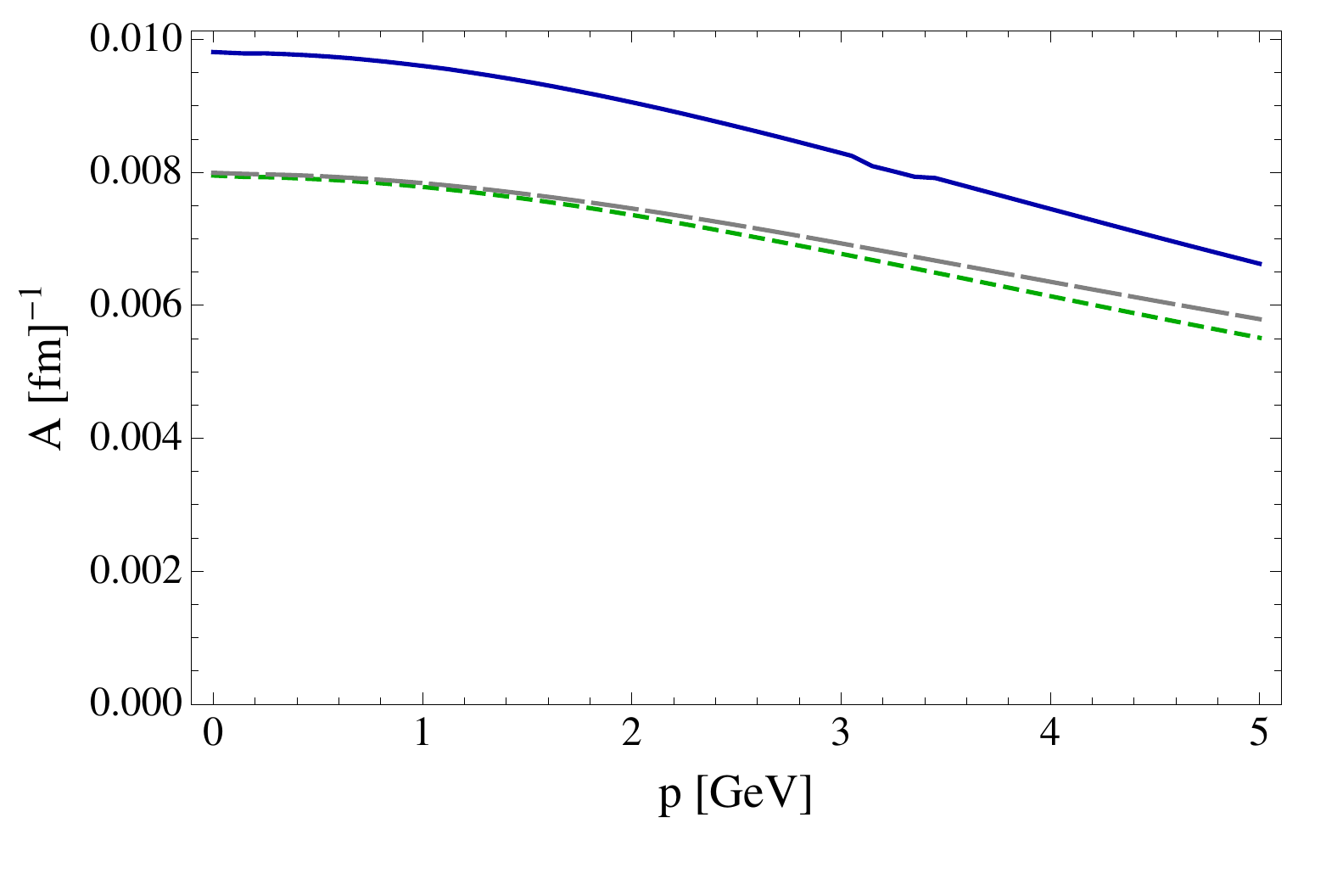}
\includegraphics[scale=.48]{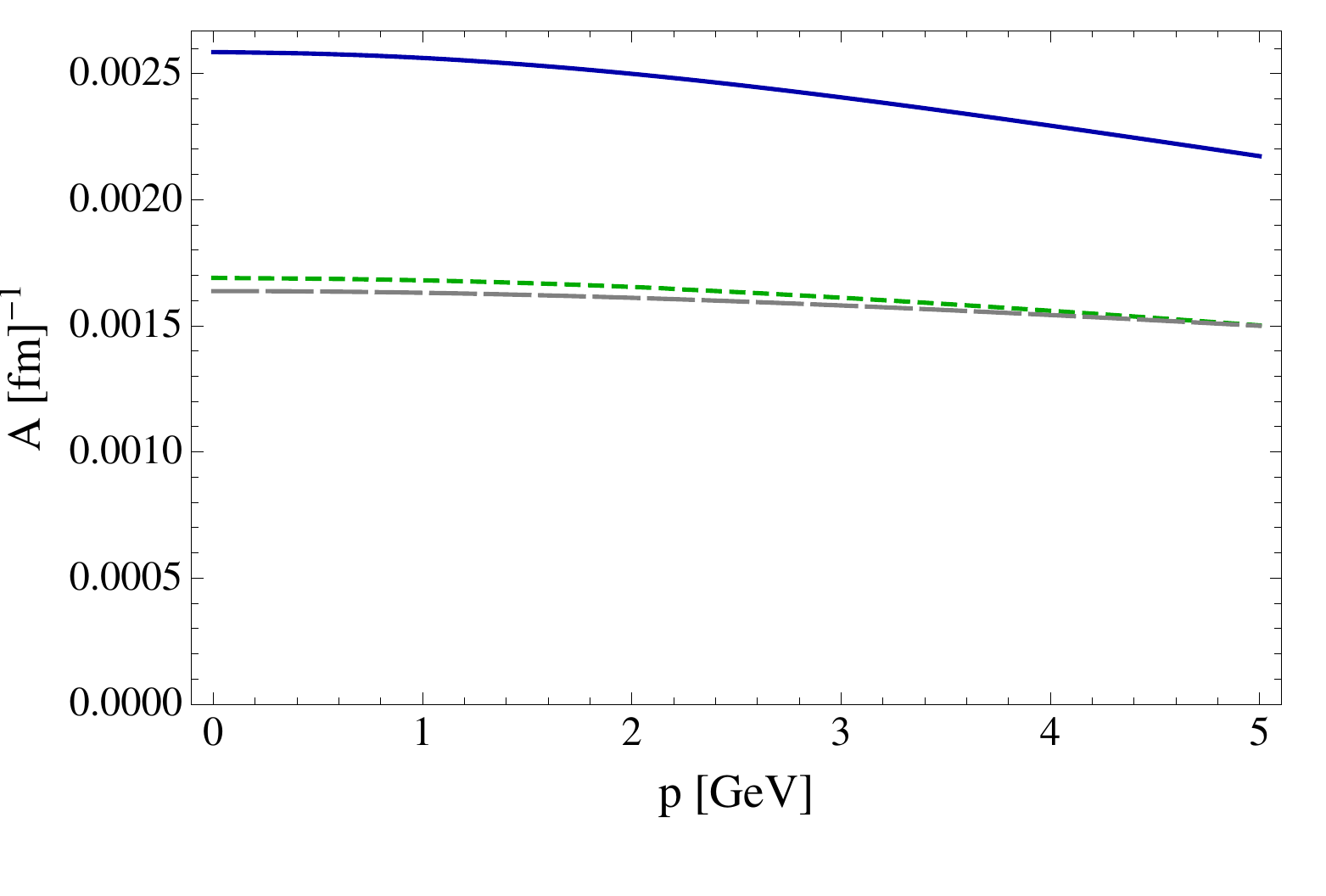}
\\
\includegraphics[scale=.48]{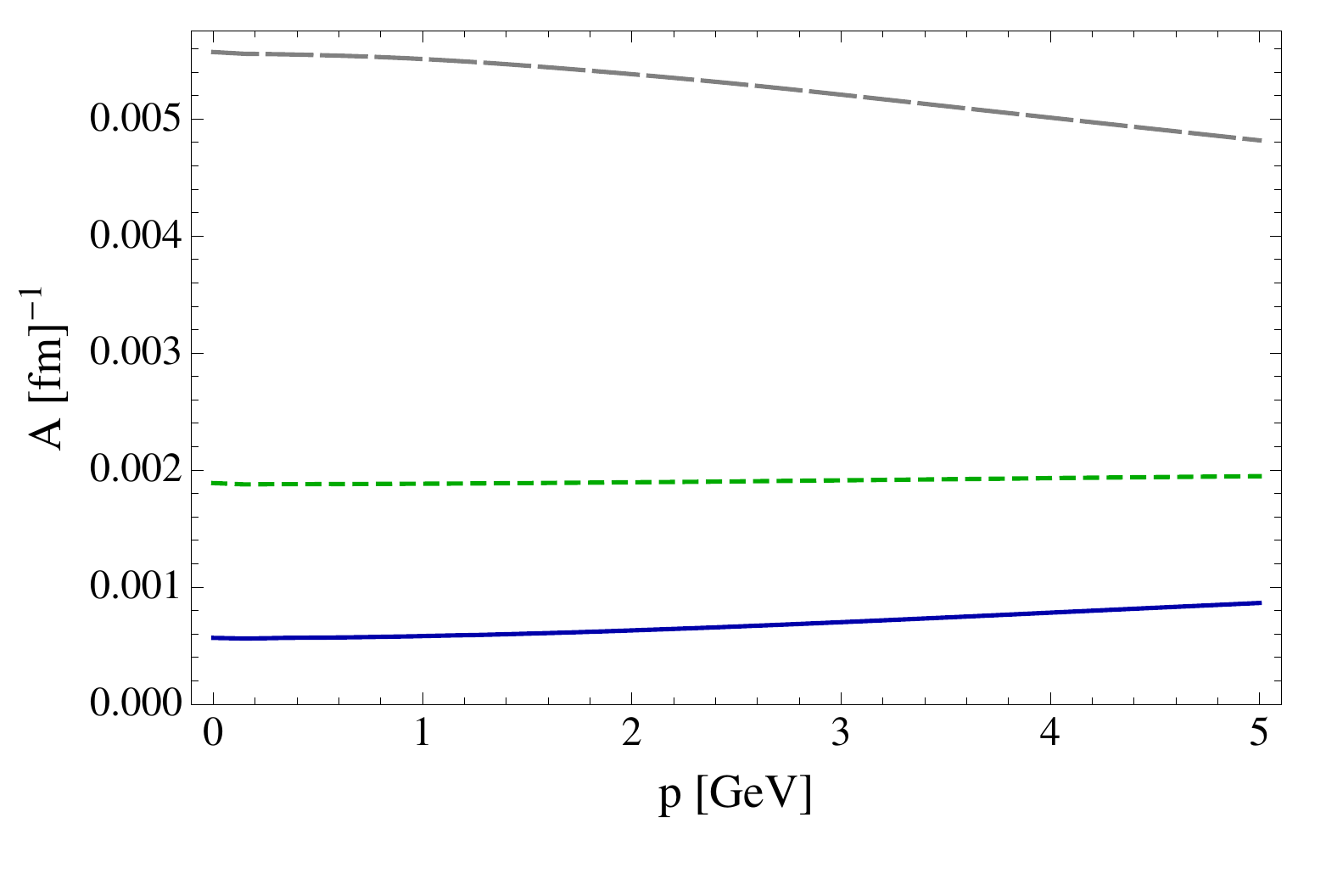}
\includegraphics[scale=.48]{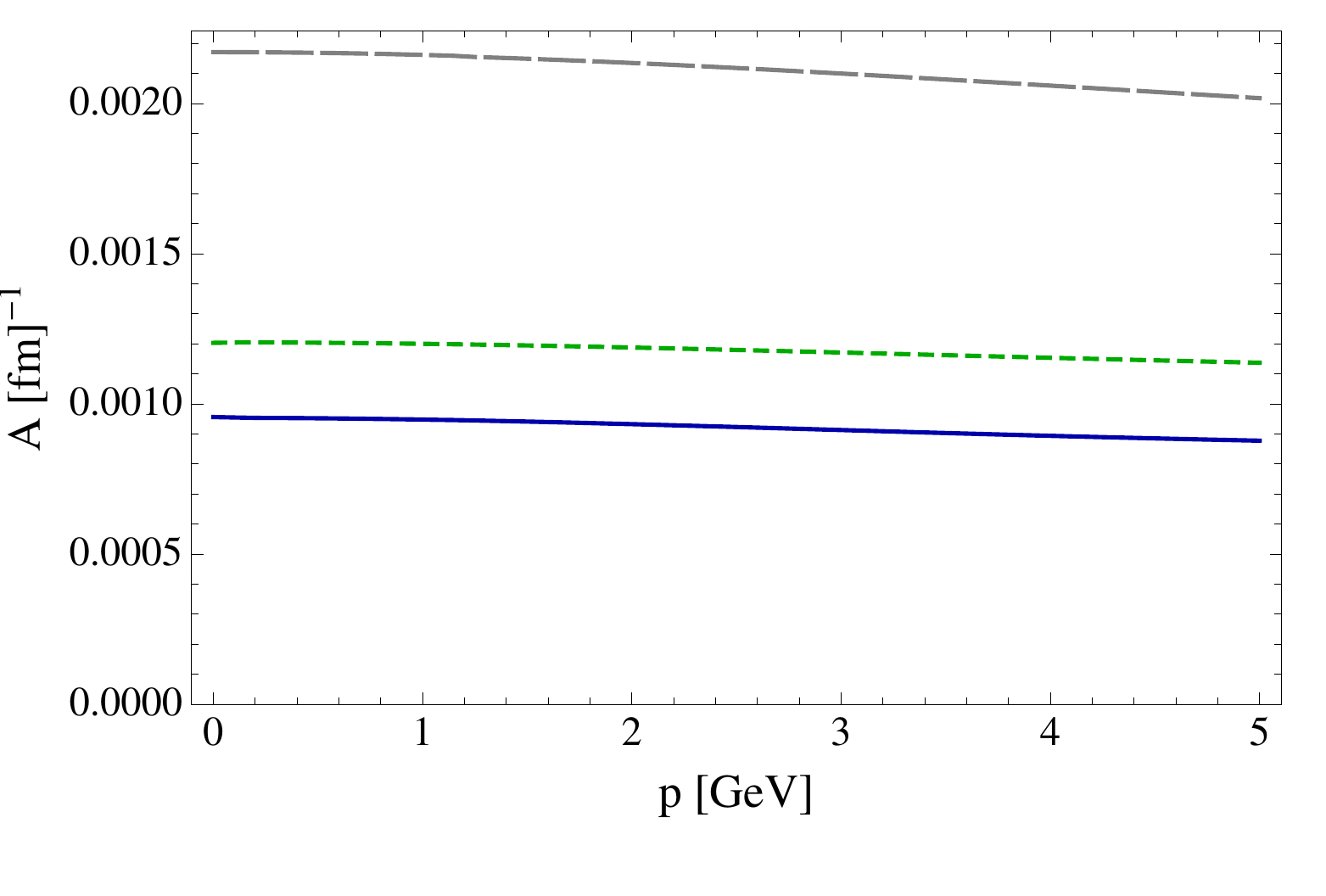}\\
\includegraphics[scale=.48]{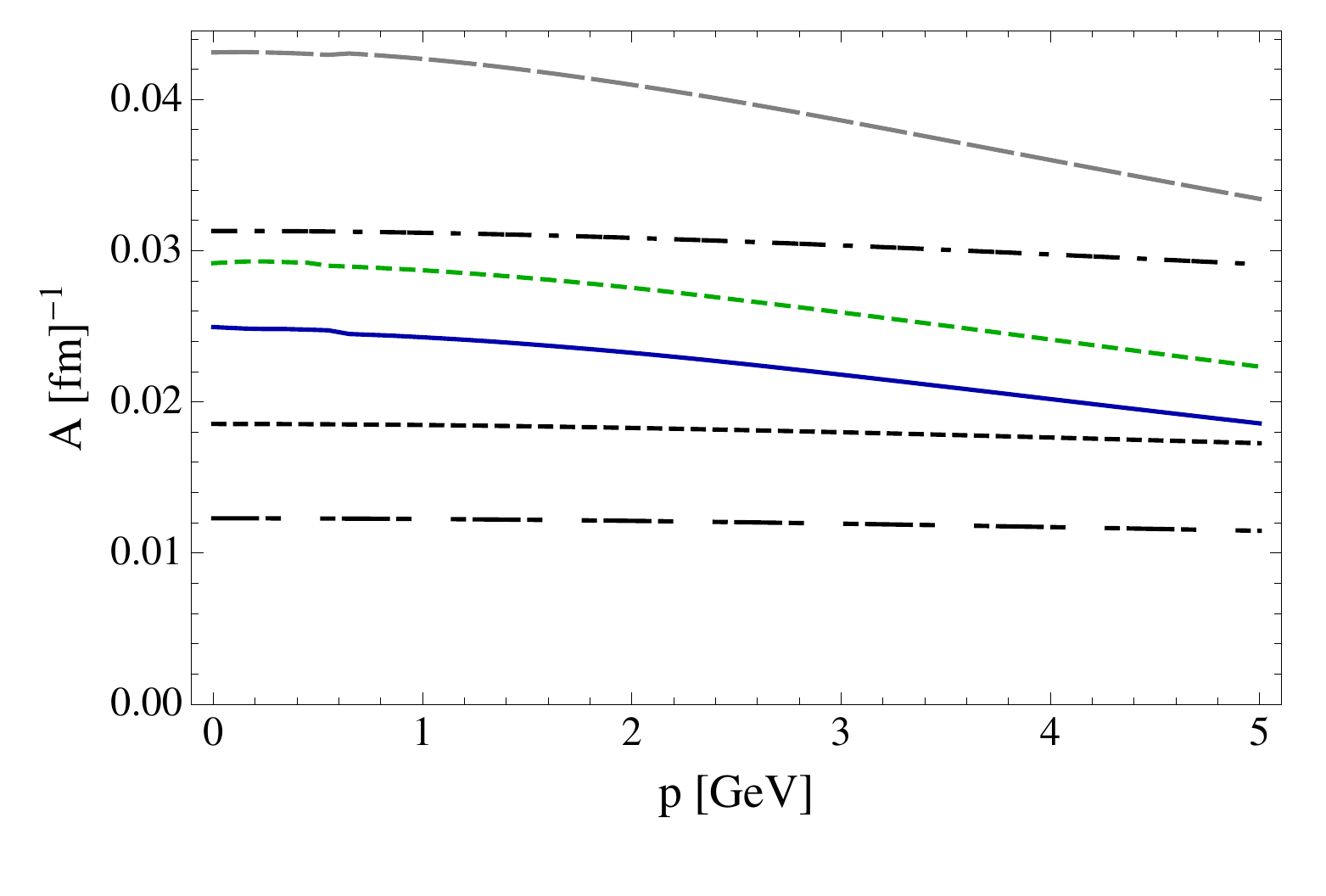}
\includegraphics[scale=.48]{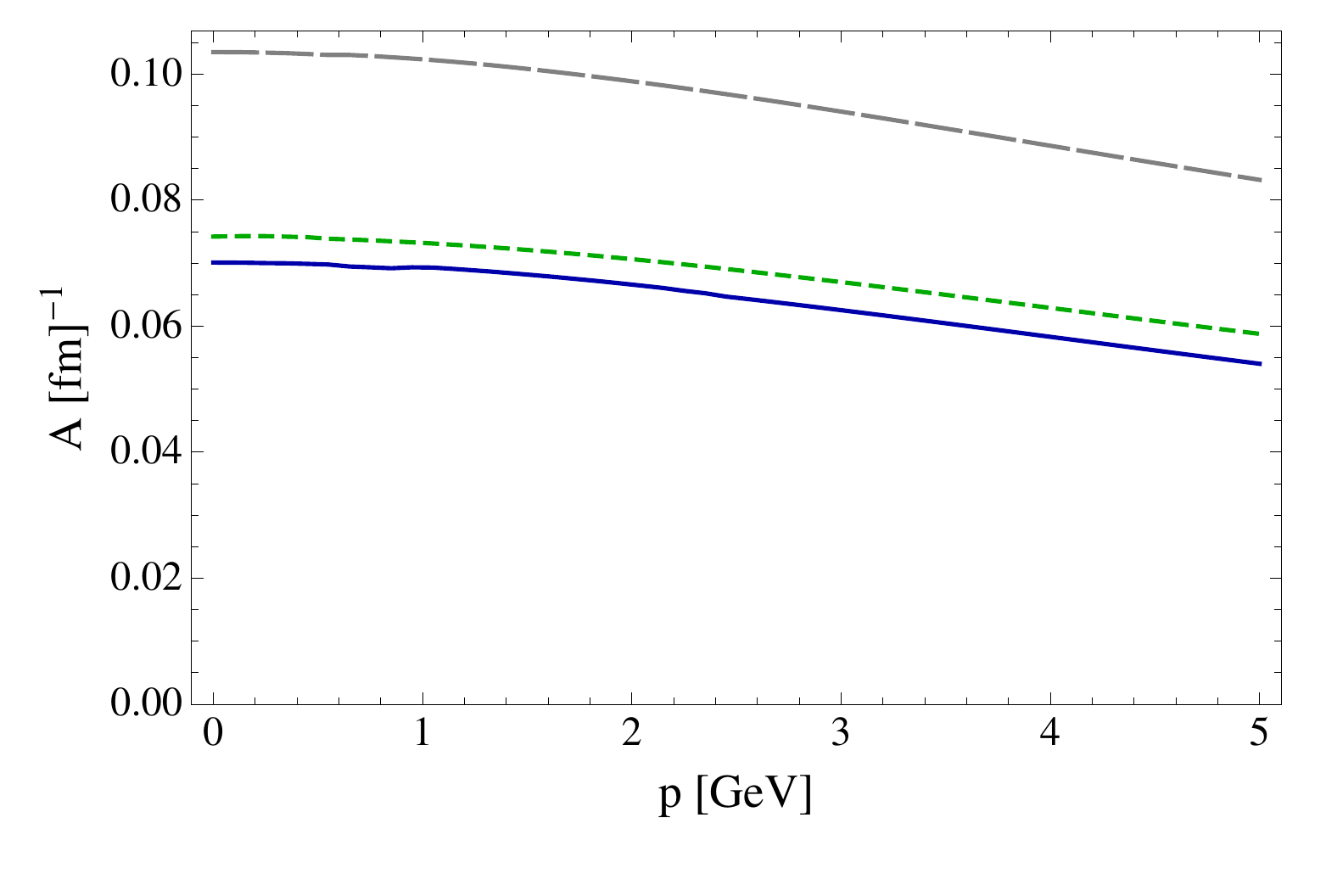}
\caption{Same as Fig.~\ref{fig_gamc-u}, but for bottom quarks.}
\label{fig_gamb-u}
\end{figure*}

The results for $\gamma_c$ employing the full $\tilde{T}$-matrix with the
$U$-potential are summarized in Fig.~\ref{fig_gamc-u}. At low temperatures
the sextet channel produces a stronger drag coefficient 
($\gamma_c=0.027$\,fm$^{-1}$) than the the triplet channel 
($\gamma_c=0.018$\,fm$^{-1}$), while the situation reverses at higher
temperature. The former is due to the larger color degeneracy in the sextet
together with the fact that the triplet resonance is still slightly below
threshold at 1.2\,$T_c$. As it moves closer to threshold, more continuum
strength becomes available for $Q$-$g$ scattering. The 15-plet gains 
little ground due to its degeneracy;  its contribution remains quite small
relative to the attractive channels. In the $P$-waves, the triplet
dominates at all temperatures considered, while 6- and 15-plet are quite
comparable. Note the small temperature dependence in the triplet (also
for the sextet $S$-wave): the large increase in scattering partners 
from the medium (thermal gluon density) is essentially compensated by
the decrease in interaction strength due to color screening.  
A qualitatively similar analysis holds for the $b$-quark relaxation
rates ($\gamma_b$) compiled in Fig~\ref{fig_gamb-u}.
\begin{figure*}[!tp]
\includegraphics[scale=.48]{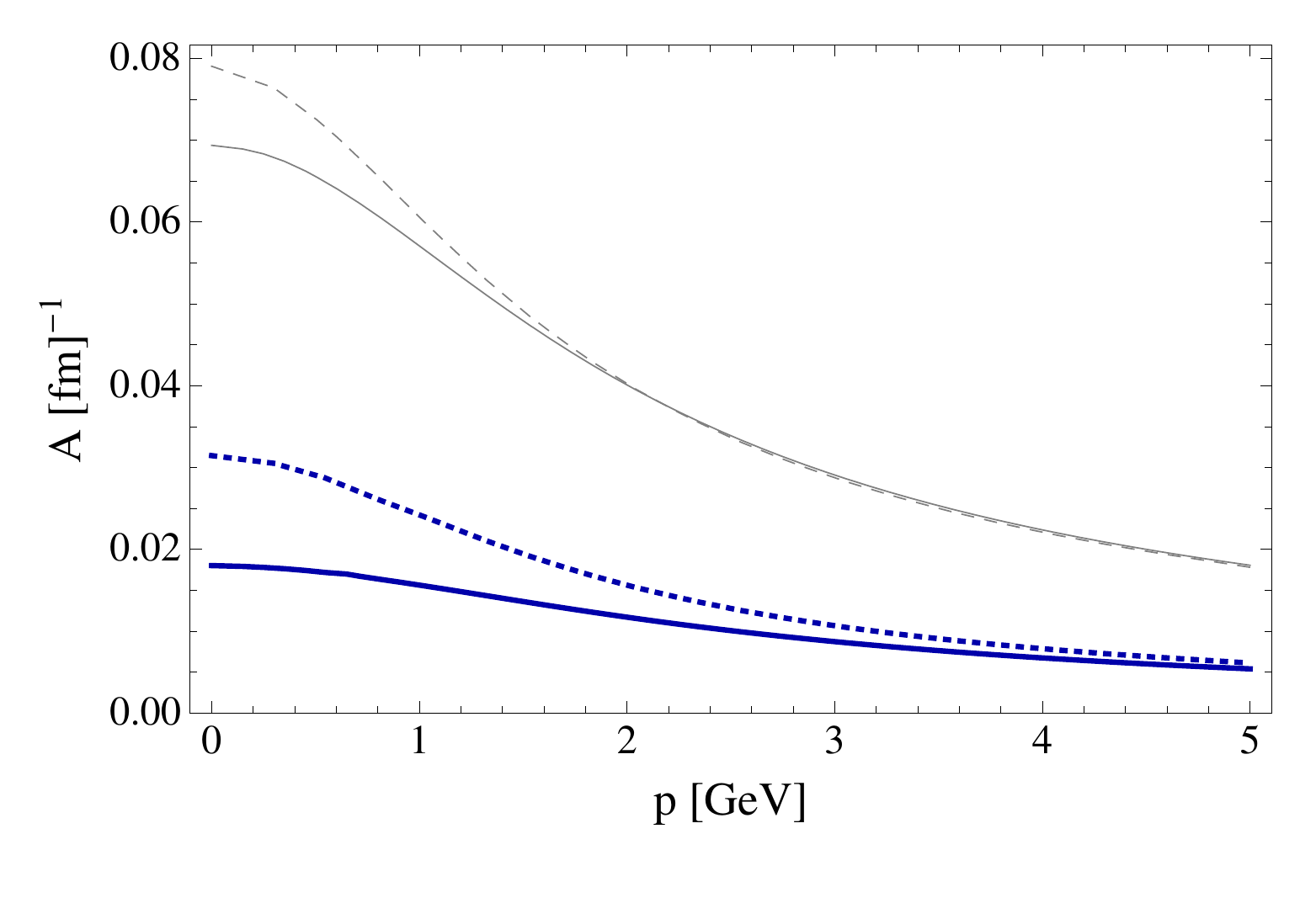}
\includegraphics[scale=.48]{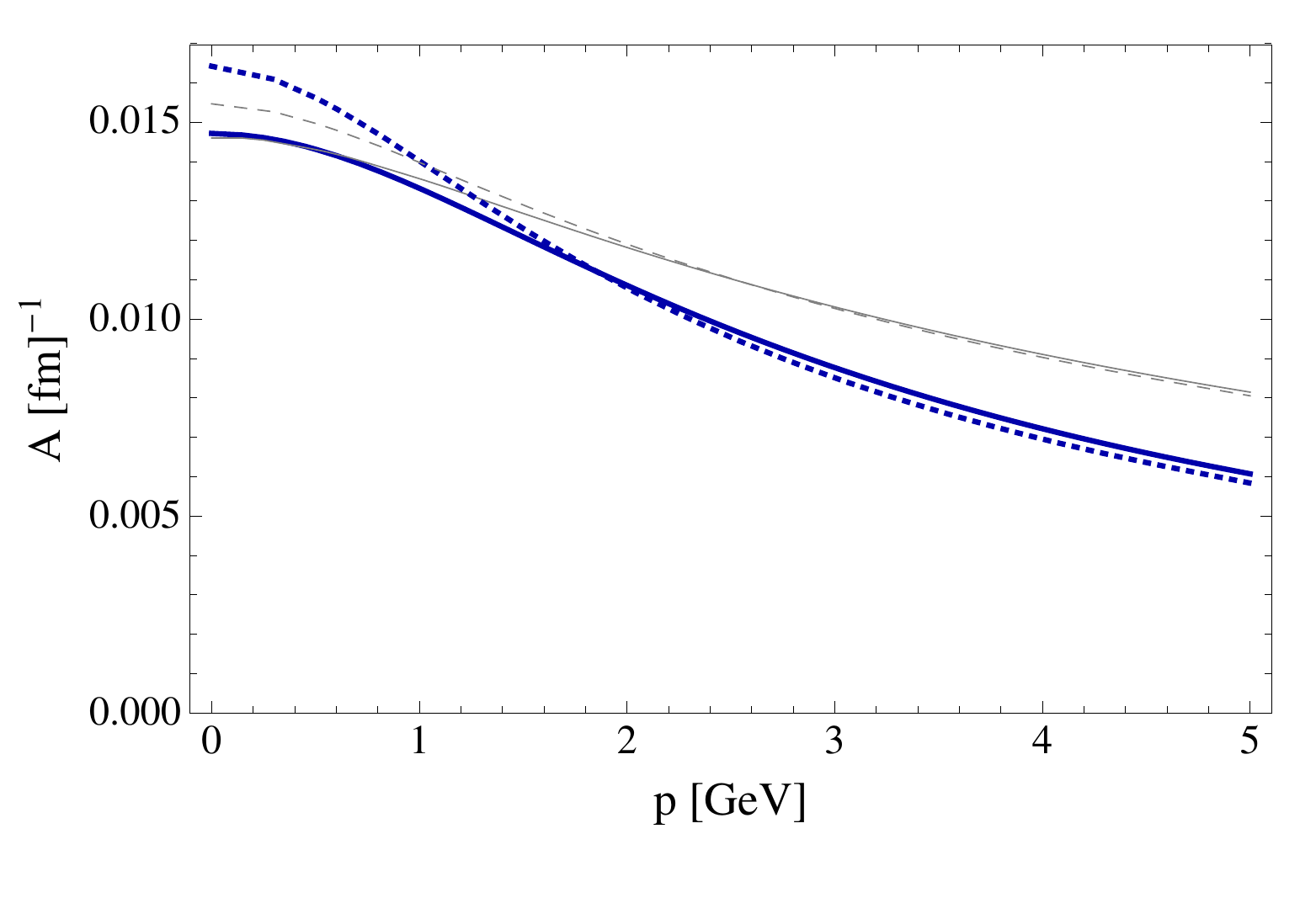}\\
\includegraphics[scale=.48]{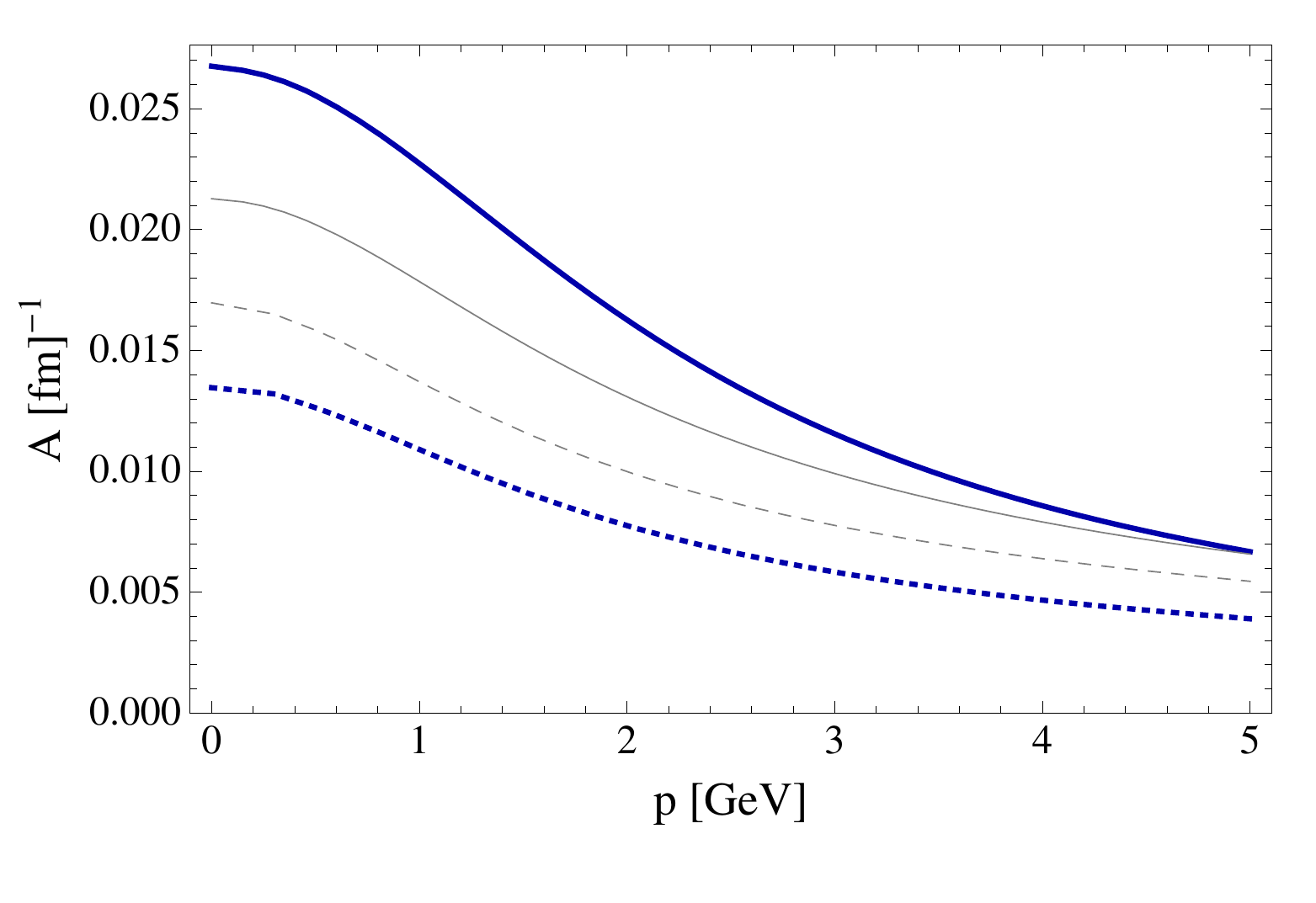}
\includegraphics[scale=.48]{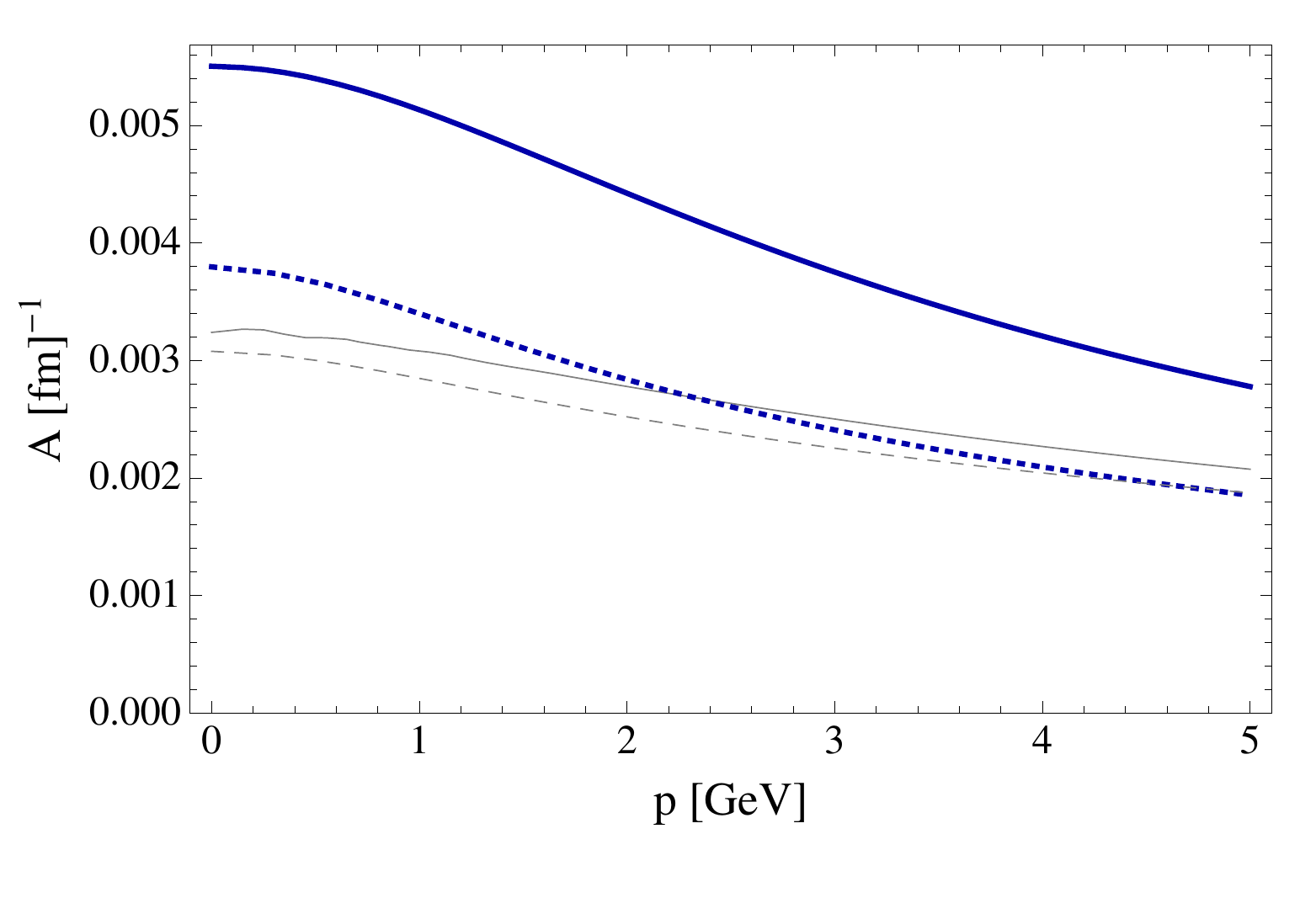}\\
\includegraphics[scale=.48]{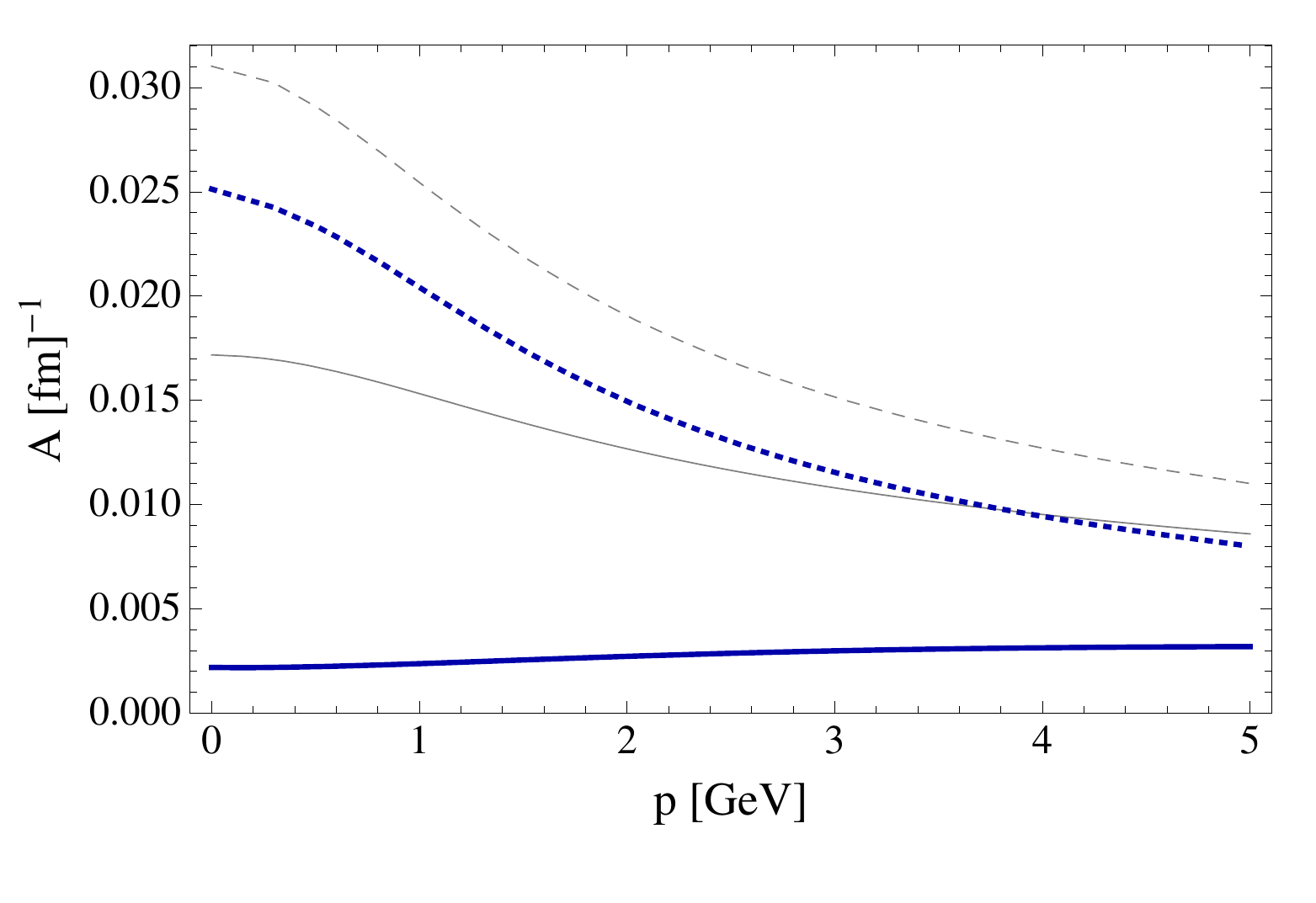}
\includegraphics[scale=.48]{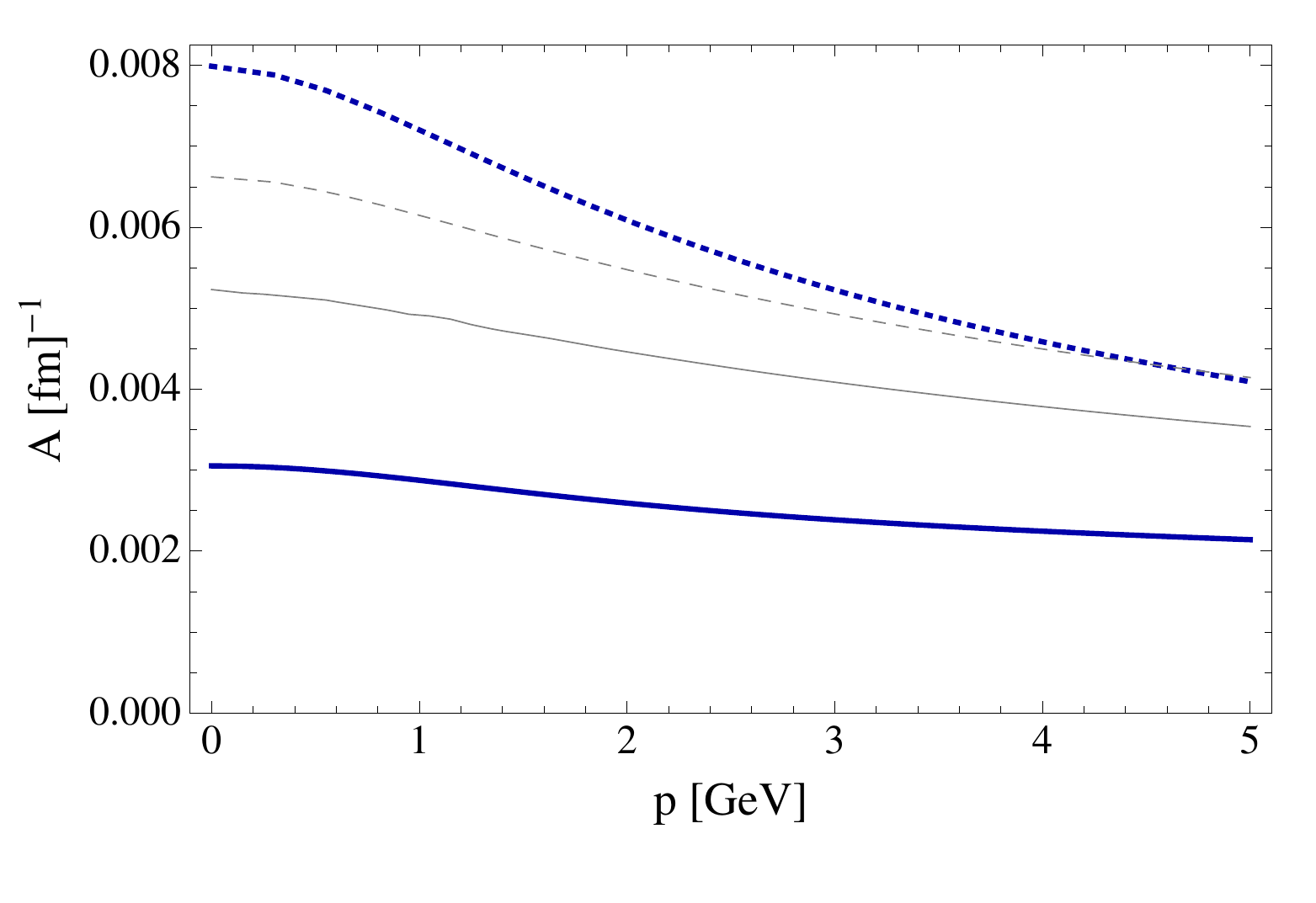}\\
\includegraphics[scale=.48]{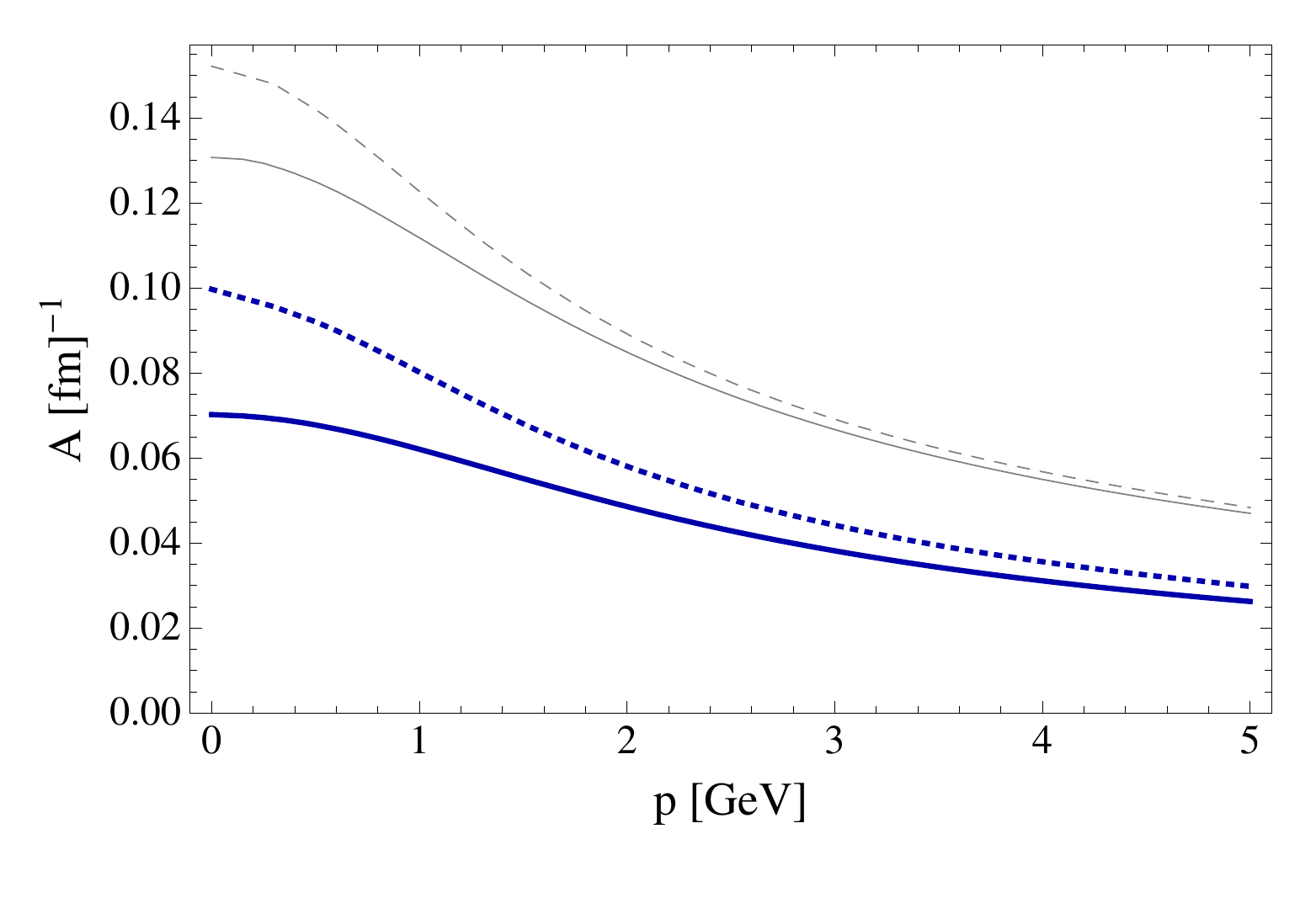}
\caption{Drag coefficients calculated with only the Coulomb part of the
in-medium $U$ potential (dashed/dotted lines), compared to the full
potential (solid lines), for charm-gluon interactions at 1.2\,$T_c$
(thick blue lines) and 2.0\,$T_c$ (thin gray lines). (Color online.)}
\label{fig_coul}
\end{figure*}

The lower right panel in Fig.~\ref{fig_gamc-u} summarizes the total 
$c$-quark relaxation rates in a QGP with 2+1 flavors as obtained from
our elastic in-medium $T$-matrices, by adding the here calculated 
gluon contributions to the $u$, $d$ and $s$ anti-/quark contributions
calculated in Ref.~\cite{Riek:2010fk} (lower left panel of Fig.~23
in there). Not unexpectedly, replacing the perturbative gluon part
in there with the nonperturbative $\tilde{T}$-matrix, the total drag 
coefficient increases by about 25\% at low momenta (somewhat more at 
1.2\,$T_c$ and somewhat less at 2\,$T_c$). However, at charm-quark
momenta of about 5\,GeV (and above), we find a slight decrease of the
total $A$. The reason is that the perturbative gluon contribution in
Ref.~\cite{vanHees:2004gq,Riek:2010fk} has been calculated with  
$\alpha_s=0.4$, while the lQCD free-energy fits result in 
$\alpha_s\simeq0.3$ for the Coulomb part (which dominates at high
momenta). This reiterates that our calculations approach the
leading-order pQCD limit at high momenta, i.e., the resummation effects
cease. In particular, it demonstrates that the use of a schematic
momentum-independent $K$-factor to upscale perturbative results in the
low-momentum regime will lead to incorrect results at high momentum.
A proper dynamic treatment of the nonperturbative effects is thus mandatory
to obtain a realistic momentum dependence of the HQ transport coefficient 
in the QGP. In a sense, this is a 
reflection of the asymptotic freedom property of QCD, with a dynamical
generation of nonperturbative effects in the low-energy domain.
Again, similar considerations apply in the bottom sector, only that
the momentum scale for recovering perturbative results is augmented
relative to the charm case by roughly the mass ratio of $m_b/m_c\simeq 3$.

\begin{figure*}[!t]
\includegraphics{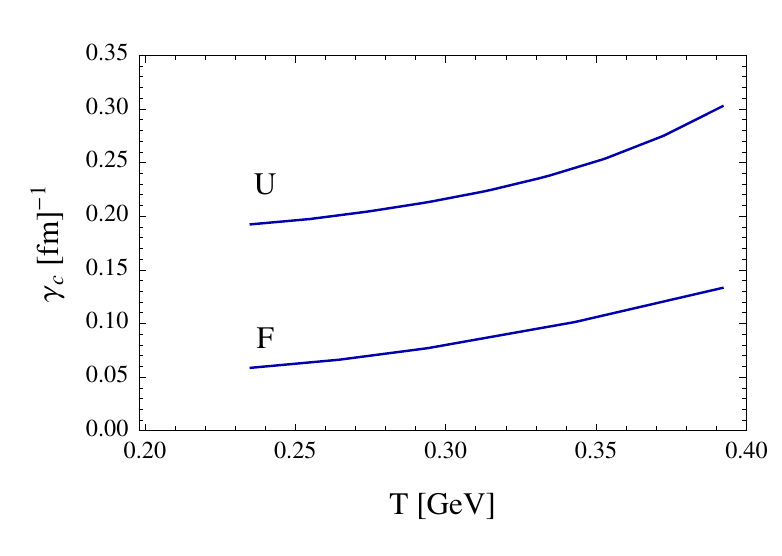}
\includegraphics{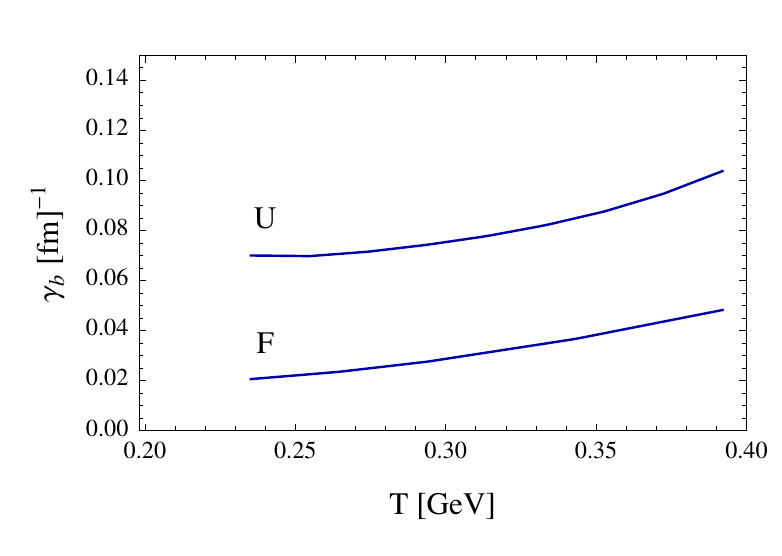}\\
\includegraphics{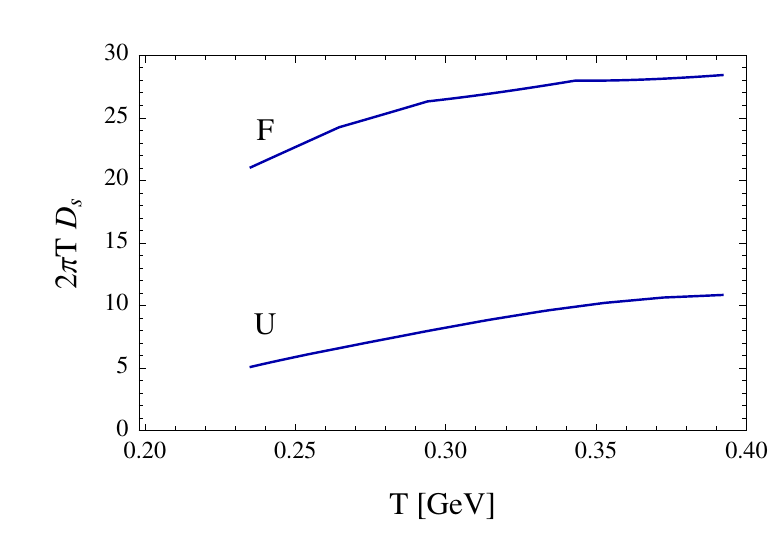}
\includegraphics{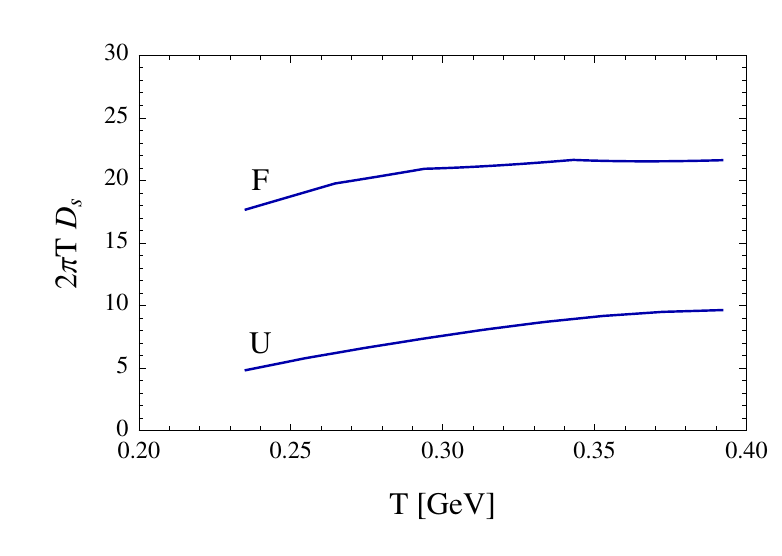}
\caption{Charm (left) and bottom (right) relaxation rates (upper) and
diffusion constants (lower) using $U$ and $F$ as a potential.}
\label{fig_transport}
\end{figure*}

We also examine the drag coefficient computed from utilizing the Coulomb-only 
potential for charm-gluon interactions, cf.~Fig.~\ref{fig_coul}. First, 
all channels are subject to
an overall kinematic effect due to the smaller in-medium charm-quark mass,
by about 20(10)\% at 1.2(2)\,$T_c$, which increases the thermal relaxation
rate accordingly. Additional effects arise dynamically due to changes in 
the $\tilde{T}$ matrix and thus vary depending on the channel under 
consideration. In the color-$ 3$ $S$-wave, there is a moderate  
enhancement of the Coulomb-only over the full potential for 1.2\,$T_c$
at small momenta which is partly due to the bound state (or resonance) 
being located closer to the 2-particle threshold owing to less binding 
in the absence of the string term; this provides more 
$\tilde{T}$-matrix strength in the scattering regime. At 2\,$T_c$ this
effect is no longer operative (the small enhancement is compatible with 
the kinematic effect alone). In the 15-plet, the enhancement of the  
Coulomb-only case over the full potential is more pronounced, especially 
at low temperature, due to the absence of the compensation between
attractive string and repulsive Coulomb terms.
In the sextet, on the other hand, one finds an overall suppression effect: 
despite the smaller $c$-quark mass, the large loss in near-threshold 
$\tilde{T}$-matrix strength due to the absence of the
string term leads to a reduction of the relaxation rate. 
The total HQ relaxation rate from interactions with thermal gluons 
increases in the Coulomb-only case relative to the full calculations
by $\sim$40(15)\% at 1.2(2.0)\,$T_c$, i.e., the nonperturbative effect 
fades away toward higher temperatures, as expected. At 1.2\,$T_c$, the
suppression of the $S$-wave sextet (factor of 0.5) is more than compensated 
by a 70\% increase in the triplet and the large enhancement of 
the 15-plet (factor of $\sim$10). At 2\,$T_c$, the effects of the missing 
string term essentially compensate each other between the different 
channels, leaving a net increase of $\gamma_c$ which is close to the 
expected mass scaling. At high momenta, where both mass and nonperturbative 
effects are expected to become suppressed, we indeed observe a general 
tendency of convergence of the full and the Coulomb-only potential results.  

We finally relate the relaxation rate at zero momentum to the spatial 
diffusion coefficients by
\begin{equation}
D_s=\frac{T}{m_Q\gamma_Q} \ .
\end{equation}
Note that this quantity essentially divides out the kinematic ``delay"
of $\sim$$T/m_Q$ of the relaxation rate for a massive particle. Thus, 
in first approximation
one may expect $D_s$ to be independent of the HQ mass. This is one of the
main reasons why this quantity, after scaling by temperature to a
dimensionless quantity, has been used as an indicator of a general
transport parameter of the QCD medium, namely the ratio of viscosity to 
entropy-density, $\eta/s\simeq c T D_s$. The coefficient $c$ typically 
ranges from 1/5 to 1/2 in weakly and strongly coupled plasmas, respectively. 
In Fig.~\ref{fig_transport} we summarize the temperature dependence of
$\gamma_Q$ (upper panels) and $D_s$ (lower panels) from our calculations
for charm (left panels) and bottom quarks (right panels), using both
free and internal energies for the input potentials. For the relaxation
rate, we roughly find the expected mass scaling by a factor of 
$m_c/m_b$ when going from charm to bottom, for both $F$ and $U$ potentials.
For charm quarks the difference between the potentials translates into
a factor of 4(2.5) increase in the relaxation rate at low (high) 
temperatures. This is somewhat less pronounced for bottom.
The anticipated independence of $D_s$ on mass holds well for the
$U$ potential (within ca.~10\%), while for the $F$ potential the
deviations are somewhat larger (ca.~20-25\%). If $TD_s$ is indeed
proportional to $\eta/s$, all results are suggestive for a shallow 
minimum toward the critical temperature, which is nontrivial since
it implies a marked increase in interaction strength with decreasing
temperature. Quantitatively, our newly included nonperturbative treatment
for HQ-gluon scattering reduces $D_s$ by about 25\% relative to
our previous results~\cite{Riek:2010fk}; for the $U$-potential both 
charm and bottom give $TD_s\simeq0.8$ at the lowest temperature, which 
would imply $\eta/s\simeq 0.16-0.4$.


\section{Conclusion}
\label{sec_concl}
In this paper we have extended a previous in-medium $T$-matrix approach 
for elastic heavy-quark scattering off light anti-/quarks in a deconfined 
plasma to the thermal gluon sector. Our main objective was the evaluation
of relativistic corrections to the static potential and a reliable matching 
of normalization factors to account for the spin-1 properties of the 
gluons. We also investigated the possibility of thermal corrections to 
the gluonic vertices and restricted the gluon polarizations to the physical 
transverse ones in the presence of a thermal mass. The diagrammatic analysis 
of Coulomb and string contributions to the potential led to relativistic 
corrections paralleling the light-quark case. Upon a ladder resummation
in the $T$-matrix equation, we found the emergence of HQ-gluon resonance 
structures in the attractive color-Coulomb channels close to the 2-particle 
threshold at QGP temperatures close to the critical one. The peak structures 
gradually dissolve with increasing temperature but an appreciable 
near-threshold enhancement persists even at 2\,$T_c$. These nonperturbative
effects in the scattering amplitudes lead to a marked enhancement of the
pertinent transport coefficient (thermal relaxation rate), by up to a factor
of 3 at low HQ momenta and close to $T_c$. At higher momenta, $p\gg m_Q$, 
the rates approach perturbative values, thus reiterating the importance
of a dynamical treatment to consistently account for the momentum
dependence of HQ transport in the QGP. We furthermore investigated the
interplay of Coulomb and string terms; by switching off the latter, a
nontrivial interplay between kinematic mass reduction, binding effects
and the lack of constructive (destructive) interference in the
attractive (repulsive) Coulomb channels was observed. These effects 
ceased with increasing temperature, signaling the fading of the string 
term.  

By combining the nonperturbative gluon contribution with previous 
calculations in the light- and strange-quark sector we found that an
overall enhancement of the low-momentum transport coefficient by about
25\% emerges, while the high-momentum part is little affected (indicating
even a slight decrease compared to the previous perturbative treatment
with larger coupling). These updates should be included in phenomenological
applications to heavy-quark observables at RHIC and LHC~\cite{He:2012df}.
When coupled with a quantitatively constrained bulk evolution model
(e.g., relativistic hydrodynamics), the updated input rates can serve
as a rather complete baseline for the contribution of elastic interactions 
in the HQ diffusion process. We recall that at low momenta these are the
only interactions contributing to the transport coefficient. When coupled 
with a realistic hadronization description and subsequent hadronic 
diffusion in a heavy-ion reaction, a detailed comparison with experiment 
can then reveal remaining shortcomings in the description, most notably 
contributions from radiative processes which are expected to become 
relevant at (much) higher HQ momenta. Work in these directions is
in progress.  

\acknowledgments
This work has been supported by the U.S. National Science Foundation
(NSF) grant no.~PHY-0969394 and by the A.-v.-Humboldt Foundation.

\bibliography{bibtex/mybib}{}
\bibliographystyle{revtex}
\end{document}